# SELF-ORGANIZED SOCIOPOLITICAL INTERACTIONS AS THE BEST WAY TO ACHIEVE ORGANIZED PATTERNS IN HUMAN SOCIAL SYSTEMS: ARGUMENTS AGAINST THE TOP-DOWN CONTROL OF CLASSICAL POLITICAL REGIMES

NATHALIE ELAINE MEZZA-GARCIA

“Self-organized sociopolitical interactions as the best way to achieve organized patterns in human social systems: Arguments against the top-down control of classical political regimes”

Monograph

Presented as a requirement to obtain the title of

Political Scientist

Submitted to the Political Science and Government School

Universidad Colegio Mayor de Nuestra Señora del Rosario

By:

Nathalie Elaine Mezza-Garcia

Directed by:

Carlos Eduardo Maldonado Castañeda

1st Semester, 2013

*A Mayía, mi bisabuelita.*

*To Mayía, my great-grandmother*

## ACKNOWLEDGMENTS

I sincerely thank the director of my monograph, Carlos Eduardo Maldonado, who formally welcomed me to the universe of the sciences of complexity, with all the kindness of the world. With the same kindness, he allowed me to assist and participate in his PhD seminars, where my love for knowledge and curiosity grew exponentially. The maturation of the idea with which I knocked his door for the first time could not have been possible if it wasn´t for his language and teachings. I owe the trust he placed on me during the year of the internship I did with him for the formalization of my curiosity. I gratefully thank Nelson Alfonso Gómez Cruz, from the Modeling and Simulation Laboratory of the Universidad del Rosario, for all his lessons and orientation. I thank him for his classes, for the conversations and his disinterested support during the writing of this monograph. Without him, it could not have been possible. I will be forever grateful to Carlos Eduardo Maldonado and Nelson Alfonso Gómez Cruz for everything that they have represented in my life so far. I thank Professor Beatriz Franco Cuervo for the friendly hand she held to me during many Fridays, when she listened to me and helped me clarify concepts that, without her help, would not appear here with the same clarity. I thank Professor Ruben Fontalvo, from the Universidad Simón Bolivar, in Barranquilla, Colombia, who is in great part responsible for helping me turn my dreams into goals, leading me to the man who became my thesis director some years later. I thank Professor Julio Robayo Lozano for motivating my interest in the organizational structures of public institutions, sowing what would later become this work. I thank my mother for her patience and the passion we share for loving a subject and surrendering to it with no hindrance; I thank my dad for his love and unconditional support; and to both, altogether, for all the trust they place on me and for never imposing boundaries to my ideas. On the contrary, I would be, statistically speaking, part of a normal distribution of a Gaussian bell and I would have never known complexity as a scientific problem. I thank Sergio Montoya, my Math Teacher from school, who introduced me to the colorful world of fractals, and Jose Gregorio Preciado, also my school teacher, for his love and teachings. Also, to everyone who was sending positive energy, family and friends, wanting nothing else but for me to succeed. And, with all my love, I thank Hoffy and Caffu, who were with me during the whole writing process of the monograph and that with a love, cuteness and tenderness, so natural in them, were always in charge of letting me know if it was too late and I had to go to bed, too early and I had to go to sleep or, simply, time to take a break and play…

**AGRADECIMIENTOS**

En primer lugar, quiero darle mis más sinceros agradecimientos a mi director de tesis, Carlos Eduardo Maldonado, quien *de mil amores* me abrió formalmente las puertas al mundo académico de las ciencias de la complejidad. Con la mayor amabilidad del mundo me permitió asistir e involucrarme en sus seminarios de doctorado, en los que mi amor por el conocimiento y la curiosidad crecieron exponencialmente. A su lenguaje le debo la maduración de la idea con la que toqué sus puertas por primera vez y a la confianza que depositó en mí durante el año que hice la pasantía a su lado le agradezco el haberme enseñado a formalizar la curiosidad. Le agradezco, de igual manera, muy sinceramente a Nelson Alfonso Gómez Cruz, del Laboratorio de Modelamiento y Simulación de la Universidad del Rosario, de quien tanto he aprendido. Le doy gracias por las clases, las conversaciones tan fructíferas que tuvimos y el apoyo desinteresado que siempre me brindó durante la escritura de mi monografía. Sin él, este trabajo no hubiese sido posible. Estaré por siempre agradecida con Carlos Eduardo Maldonado y Nelson Alfonso Gómez Cruz por todo lo que han representado en mi vida hasta hoy. A Beatriz Franco Cuervo, profesora de mla Facultad, le agradezco el que, como una mano amiga, durante varios viernes me regaló horas para escucharme y aclararme conceptos que, sin su ayuda, no hubiesen podido quedar plasmados en esta monografía de manera clara. Al Profesor Rubén Fontalvo, de la Universidad Simón Bolivar, en Barranquilla, mi más profundo agradecimiento, pues de no haber sido por él no habría conocido a quien tres años después se convertiría en el director de esta monografía. Al Profesor Julio Robayo Lozano, de la Universidad del Rosario, quien motivó en mí el interés por las estructuras organizacionales, sembrando gran parte de lo que luego se convertiría en este trabajo. Por supuesto, a mi mamá le agradezco por compartir la pasión que se siente al amar el conocimiento y entregarse completamente a él; a mi papá por su amor y apoyo incondicional; y a ambos, en conjunto, por haber buscado un balance entre depositar total confianza en mí y no haberle impuesto límites a mis ideas desde niña o a lo que yo era capaz de hacer. Sin ello, sería, en términos estadísticos, un individuo que haría parte de una distribución *normal*, y no habría llegado a la complejidad como problema científico. A mis profesor de Algebra del colegio, Sergio Montoya, pues gracias a él conocí la belleza de los fractales; y mi profesor de inglés y amigo del colegio, Jose Gregorio Preciado, por todo su cariño y enseñanzas. También, a todos aquellos y aquellas, familia, amigos y amigas, que me apoyaron con energía positiva. Por último, con mucho amor reconozco a Hoffy y a Caffu, quienes estuvieron conmigo durante todo el proceso de redacción y que con el cariño y la ternura que sólo ellas pueden tener me avisaban si ya era muy tarde y tenía que levantarme, si ya era temprano y tenía que ir a dormir, o, simplemente, si era tiempo de tomar un descanso y jugar.

*"It is impossible to dedicate and study politics in a serious manner if there is not a solid idea according to the best of the progress in knowledge about life, in general, and living systems"*

Carlos Eduardo Maldonado

"*To be governed is to be watched, inspected, spied upon, directed, law-driven, numbered, regulated, enrolled, indoctrinated, preached at, controlled, checked, | estimated, valued, censured, commanded, by creatures who have neither the right nor the wisdom nor the virtue to do so*"

Pierre Joseph Proudhon

**TABLE OF CONTENTS**

**LIST OF FIGURES AND TABLES**

# INTRODUCTION

Human social systems are complex systems. They are open, decentralized, dynamic and adaptive. They are conformed by a large number of interconnected and interdependent elements; unpredictability and uncertainty are their distinctive features; they present evolutionary dynamics and, as many other complex systems, they have an intrinsic tendency to self-organize their interactions. Examples of self-organized interactions in human social systems are economic exchanges, human mating and friendship bonds, which are not controlled, imposed or commanded by any central or top-down authority. Instead, the latter follow local and individual decisions.

Despite the tendency of human social systems towards self-organization, historically there have existed specific political systems provided with the function of organizing human interactions by means of particular institutions and organizations. These institutions and organizations, referred to as *political regimes,* are endowed with characteristics and properties that oppose those of human social systems because they are rigid, non-adaptive, non-evolvable and, most importantly, they operate with top-down control. Additionally, political regimes emphasize in normative principles and physical coercion, which frames and inhibits, not only the self-organization of human social systems, but also their complexity. It is commonly argued that in the absence of normative institutions that impose order upon human social systems, it would be impossible to deal with their vastness and diversity, leaving space for disorder and disruption. Yet, this monograph looks to provide arguments against this assumption on the basis of the property of self-organization that complex systems present when they are not being constrained by an external entity.

Thereby, the thesis of this monograph is that since human social systems are complex systems, the self-organization of their sociopolitical interactions is the best way in which organized global patterns can emerge. This, instead of being organized in a top-down fashion by political regimes that in order to accomplish their function of organizing the latter use coercive mechanisms of control exerted by means of hierarchical structural arrangements. Global pattern refer to the emergent structure of local interactions and dynamics.

The main goal of the monograph will be to propose the self-organization of human sociopolitical interactions as the best way in which human social systems can make organized patterns to emerge, given that organizing them by means of top-down control is not suitable for properly addressing their complexity. This will be achieved by evidencing the complexity of human social systems and contrasting it with the characteristics of classical political regimes; presenting the plausibility of self-organized human social systems as preferable to top-down controlled; explaining how could the theory of self-organization in living systems can provide the theoretical background for sociopolitical self-organization in human social systems, given that they are complex systems and present life-like properties; showing that historically classical political regimes have tried to organize human social systems by means of top-down control; summarizing the disadvantages of organizing human social systems in a top-down fashion; and introducing what could be the emergent global pattern of sociopolitical self-organization.

Apart from humans, the tendency to self-organize is present in many other social animals, such as harvester ants, bacteria colonies, fish schools, flocks of birds and bees swarms. All these manage to harmonically (self) organize without recurring to any top-down structure, regime, leader or governor. Similar examples of self-organization in complex systems are the neural functioning of the brain and galaxy formation. Also, embryonic morphogenetic development, where no part of the system commands in which type of tissue cells must become. They all are complex systems and completely deprived of any central or top-down control -and yet, they produce very organized patterns. In fact, biological and physical self-organization have proven to be optimal in the production of order in large complex systems. The 13.7 billions of years of history of the universe are prof of it.

The background idea of the monograph is that if top-down control were the best way to organize human social systems, at present the dramatic levels of inequality around the world would be very different and there would not exist any individual or group acting *outside* the boundaries of political regimes. It is assumed that everyone would be satisfied with their capabilities in life, after being organized in a top-down fashion.

Although political systems have varied throughout history, the political regimes used to accomplish their function of organizing human social systems share the same organizational structure: a hierarchical pyramid with top-down power distribution, centralized control mechanisms and cause-effect expected interactions. This particular architecture is known in graph theory and classical network theory as a *tree topology*. Tree topologies as the structures of political regimes are hindrances that do now allow human social systems to self-organize, mainly, because trees are incapable of reflecting the complexity of the sociopolitical interactions of human social systems, for many reasons that go beyond their structures. Just to mention a few, their systems of laws and the separation made between the political and the civil society.

Thereby, it is valid to ask why, given the complex nature of human social systems and their tendency to self-organize, political regimes constraint them when seeking their organization? Shouldn´t political regimes allow the self-organization of human sociopolitical interactions when organizing them? Is it possible that regimes complexify to in order to properly organize human social systems? That is, are regimes capable of complexifying their dynamics until becoming the mayor expression of self-organized sociopolitical interactions? If the answer is yes, it is plausible to think about which dynamics, structures and rules in political regimes could mirror the complex nature of human social systems and enable their intrinsic tendency to self-organize or if, inevitably, political regimes will disappear in the process. Because, anyway, if there is an intrinsic tendency towards self-organization in human social systems in the absence of normative institutions and top-down control, are political regimes really necessary? This monograph approaches these inquiries to criticize the top-down organization of human social systems by means of political regimes that work with top-down control.

The monograph starts from recognizing how the sciences of complexity and their knowledge about complex systems can support the self-organization of human sociopolitical interactions because, following Thomas Kuhn´s book *The Structure of Scientific Revolutions*, we are in the middle of a scientific revolution.[1] Complexity is changing the way of understanding certain problems in science, after the many limitations that classical models have when it comes to dealing with uncertainty, unpredictability, irreversibility, non-determinism and nonlinearity. Accelerating technological advances are making more difficult to organize human social systems by means of top-down control[2], and political regimes need to reflect these changes. Thus, transforming their dynamics, rules and topologies for the evolution of sociopolitical self-organization would be a good way to start.

[1] Compare Maldonado, Carlos Eduardo. *Complejidad: Revolución Científica y Teoría.* 2009.

The subject studied in this monograph is relevant, firstly, for the discipline of political science because it insists on the limitations of classical political regimes when organizing human social systems by means of top-down control. It is claimed here that sociopolitical self-organization is the best ways in which a harmonic organization of human social systems can be reached.

Political science is one of the social disciplines that has apprehended complexity the least. So much that this is the first work ever written that studies the implications of the topological properties of political regimes in respect to the (self) organization of human social systems, in the context of the sciences of complexity –and in contexts of classical science. An important contribution of the monograph relates to historying the architectures that the milestones of political regimes have had in the past two and a half millennia in western world, in order to illustrate how human social systems have been top-down organized. This allows visualizing a truth that was hidden behind changes regarding how rulers become so: since the Greeks there has not been any profound transformation in the way in which human social systems are organized. Secondly, this monograph is relevant for the Political Science School where it was developed because it is the first work ever written about the sciences of complexity in the School, in undergraduate and graduate level. Complexity is in great part responsible for the most advanced science that is taking place nowadays. Hence, this first undergraduate work opens the door for a whole new spectrum of knowledge in the School. Thirdly, the subject of the monograph benefits the author´s personal academic development because in her future academic life she will continue to extend the study of the complex nature of human sociopolitical interactions.

The monograph divides in five chapters. Firstly, the theoretical and conceptual framework is defined. Secondly, part of the literature revision is discussed. Thirdly, it is introduced how could sociopolitical self-organization be triggered and the relation between sociopolitical self-organization and self-organization in living organisms and systems that present life-like properties is discussed. Fourthly, it is presented that, throughout history, human social systems have been organized in a top-down fashion and some of its negative implications and disadvantages are mentioned. Finally, the monograph ends with some concluding remarks and ideas about future work in relation to the subject.

---

[2] Compare Mezza-Garcia, Nathalie. "Bio-Inspired Political Systems: Opening a Field". Proceedings of the European Conference on Complex Systems ECCS´12, Belgium, 2013.

# 1. THEORETICAL AND CONCEPTUAL FRAMEWORK

This chapter introduces the theoretical and conceptual framework of the monograph, which uses theories of political science and the sciences of complexity, but also feeds from elements of classical and complex networks theory.

## 1.1. POLITICAL SYSTEMS AS SYSTEMS THAT PROCESS INFORMATION

This monograph is interested in how political regimes organize human social systems. However, before entering into what political regimes are, it is crucial to define political systems. A *Political system* is an abstract and general concept that comprises the aggregate of decision making processes in a human social system.[3] Political systems take the inputs of human social systems (demands, resources and constraints)[4] and transform them into decisions -which are later *applied* by means of political regimes in an authoritarian fashion.[5] However, before doing so, demands compete between them, resources are mobilized and constraints are prioritized. It is claimed in this monograph that because political systems transform inputs into something else, i.e., that they are systems that process information, they can be understood as computational systems because one of the meanings of computation is information processing[6]. On the other hand, the competition of demands, the mobilization of resources and the prioritization of constraints –the computational dynamics of political systems- follow lengthy processes of discussion, conciliation and studying, in principle, in order to find the best alternative -also known as the optimal. Therefore, political systems are computational systems that look to solve optimization problems.

Democracy, Dictatorship, Monarchy, Aristocracy and Theocracy are examples of political systems. Figure 1 and figure 2 show two of the most recognized models for political systems: David Easton´s (figure 1) [7] and Jean-William Lapierre (figure 2)[8]. As it can be seen, Lapierre went farther than Easton because he tried to explain which are the internal dynamics of political systems that lead inputs to be transformed into decisions, instead of simply presenting the direction of the flow of information. Nonetheless, both approaches erroneously assume that it is possible to have complete knowledge about all the elements and interactions in political systems and that their dynamics can be explained by linear causation and direct relations between inputs and outputs.

---

[3] Compare Lapierre, Jean-William. *El Análisis de los Sistema Políticos.* 1976. p. 39.
[4] Compare Lapierre. *El Análisis de los Sistema Políticos.* p. 53.
[5] Compare Easton, David. "An Approach to the Analysis of Political Systems". *World Politics.* Vol. 9. No. 3. 1957. p. 383.
[6] Compare too Fraley, Denis J. "Computation is Process". *Ubiquity Symposium "What is Computation?".* 2010; Compare Mitchel, Melanie. *Complexity: A Guided Tour. (*2009).

**Figure 1. Model of Information processing in political systems by David Easton.**

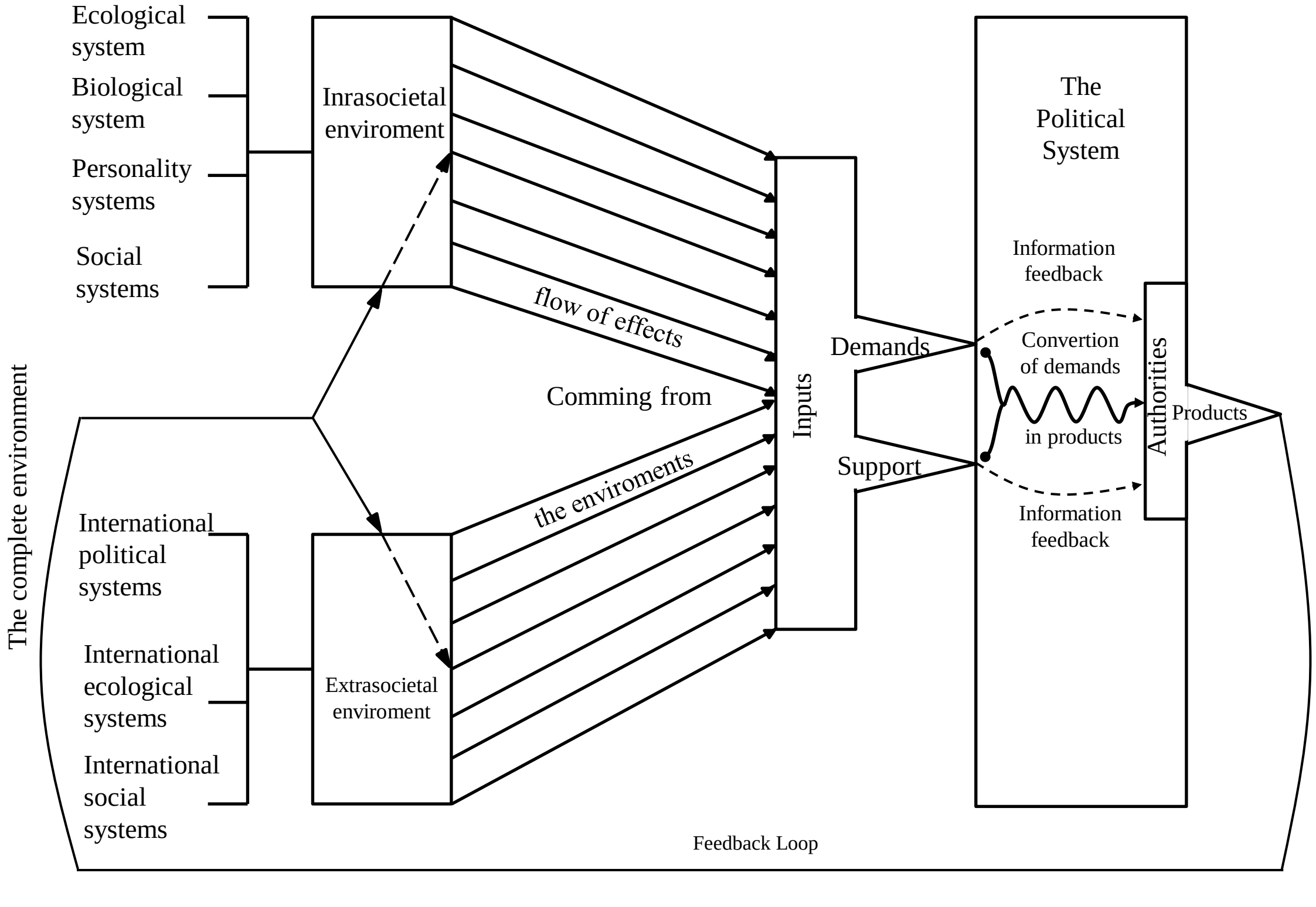

Source: Easton, David, In *Esquema para el Análisis Político,* 2006. p. 154

[7] See Easton, David. *Esquema para el Análisis Político.* 2006. p.154.
[8] See Lapierre. *El Análisis de los Sistemas Políticos.* p.49.

**Figure 2. Model of Information processing in political systems by Jean-William Lapierre .**

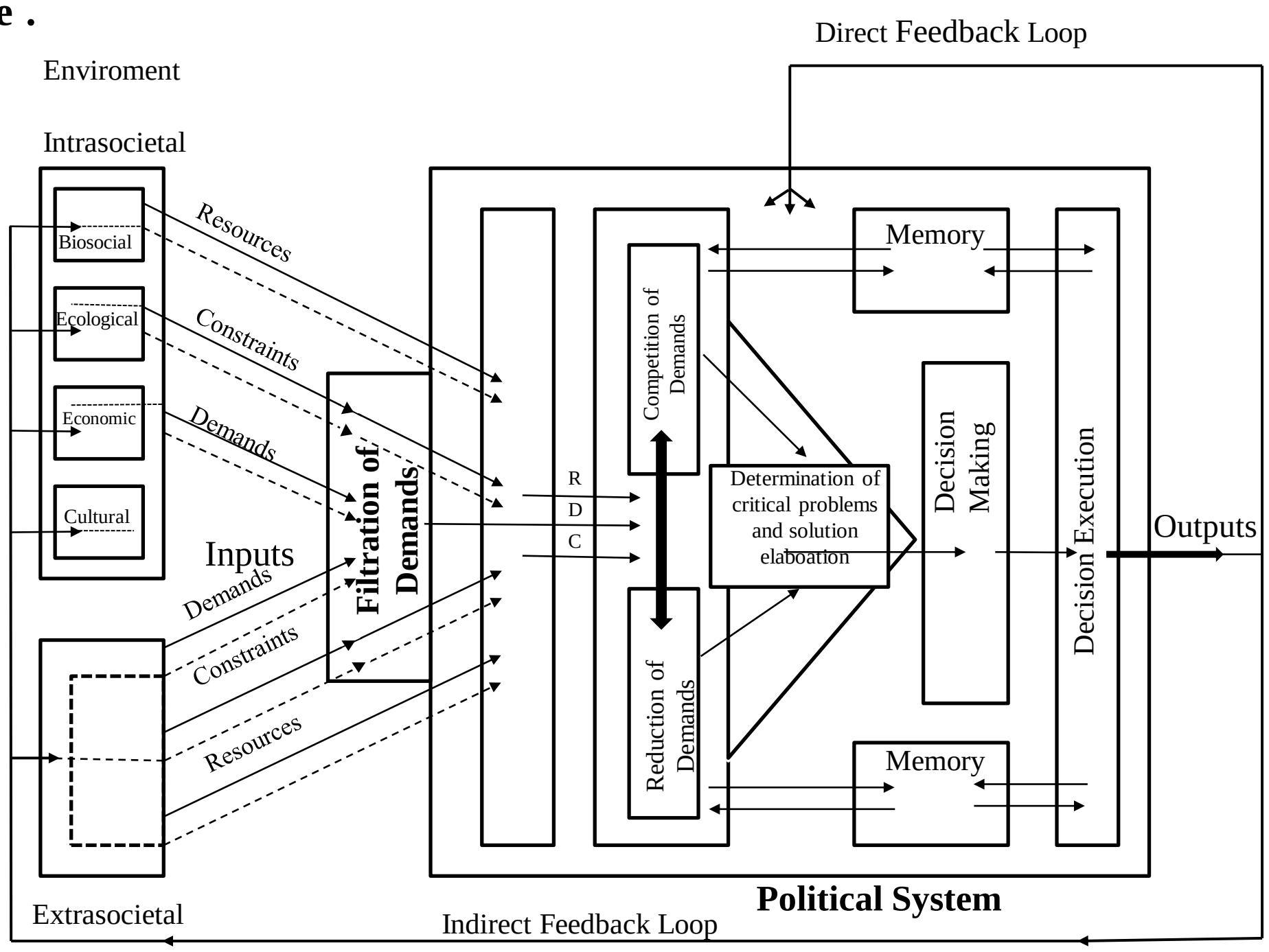

Source: Jean-William Lapierre. In *El Análisis de los Sistemas Políticos,* 1976. p.49.

One reason why political systems are conceived in this way is that the theories about political systems were developed during the heyday of General Systems Theory[9] and Cybernetics[10], both widely recognized for their interest in studying systems from holistic perspectives. It also explains why the latter models conceive political systems as if they had defined boundaries with their environments (human social systems), considering them as separated entities[11]. Nevertheless, this separation is not shared in this work since the civil society has a political dimension with political roles.

[9] Compare too Bertalanffy, Ludwig Von. *General Systems Theory: Foundations, Development, Applications.* 1969.
[10] Compare too Von Foerster, Heinz. *Las Semillas de la Cibernética.* 1991.
[11] Compare too Forgacs, David. *The Antonio Gramsci Reader. Selected Writtings.* 2000.

## 1.2. POLITICAL REGIMES

Political regimes can be understood as the last name of political systems. For instance, democratic political systems can either have presidential or parliamentary political regimes. In other words, a political regime is the organizational scaffolding of a Political system[12] or the "institutional configuration of government"[13]. It includes the set of rules for political interactions and the formal institutions that give shape to them.[14] That is, how a political system functions (its rules and dynamics) and how it is organized (its structure)[15] (figure 3). The structure is the architecture of the institutions that regimes use to implement the rules. Rules are the specification for dynamics, which can also be coded in the structures. However, the do not depend exclusively on them. In respect to human social systems, their dynamics can be predefined by the rules and structures of political regimes, but it does not mean that they are necessarily going to behave as expected.

**Figure 3. Interactions in political regimes.**

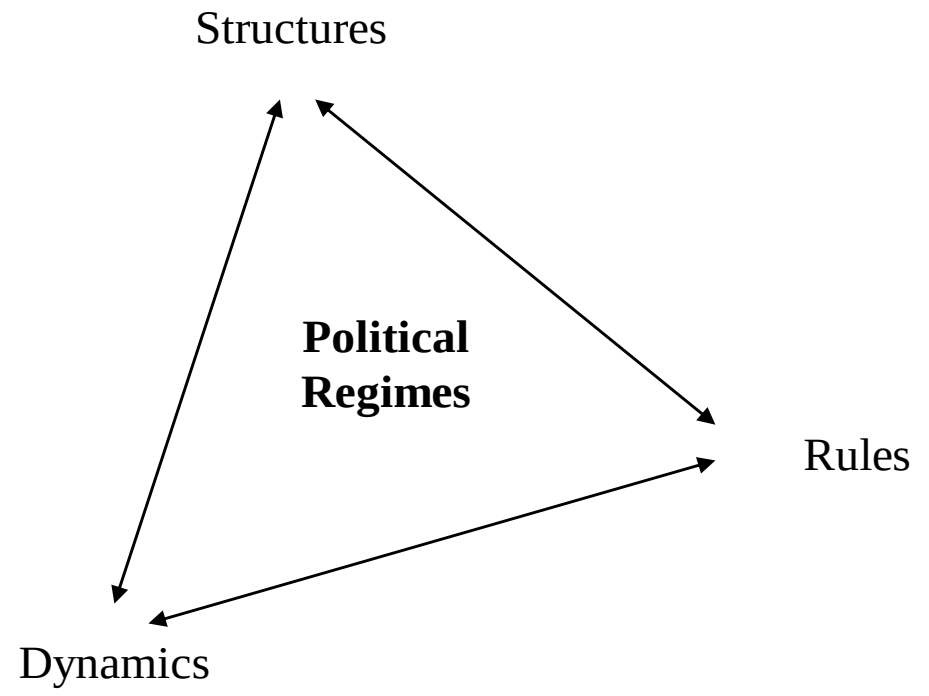

Source: Author´s own elaboration.

In graph theory, classical network theory and LAN design[16], the structure of classical political regimes is called a tree topology (figure 4). Trees developed as institutional responses to the problem of how to organize human social systems. Trees have a core channel with a node on the top. In political regimes, this node can be a king, an emperor, a president, a prime minister, a parliament or any other type of government or governor. Control channels, the backbone of the structures of trees, are in charge of the information processing of the system, which takes place in a top-down fashion. Nodes break of from the central channel and are subordinated to the node on top, in lesser or greater extent, depending on the level of the hierarchy where they are located, and to their parent node. Trees use centralized control mechanisms, top-down power distribution, linear decision-making processes and pro-equilibrium organizational structures, based on erroneous conjectures about non-complex, transparent and linear interactions in human social systems. This monograph resorted to classical network theory, specifically to LAN design, to study the implications of the tree structure and dynamics of political regimes for the organization of human social systems, since in political science there is not any theory about the topological properties of political regimes that can enlighten a proper organization of human social systems.

[12] This definition was provided [private conversations] by Professor Beatriz Franco Cuervo, who used it to define political regimes.
[13] See Kelly, Charles. *Political Science. Basic Concepts.* Electronic document.
[14] Compare Lapierre. *El Análisis de los Sistemas Políticos.*
[15] Compare Nedelcu, Paul-Iulian. "State Structure and Political Regime Structure". (2012). p. 289.
[16] Local Area Networks for connecting microelectronics-based devices.

**Figure 4. Tree topology.**

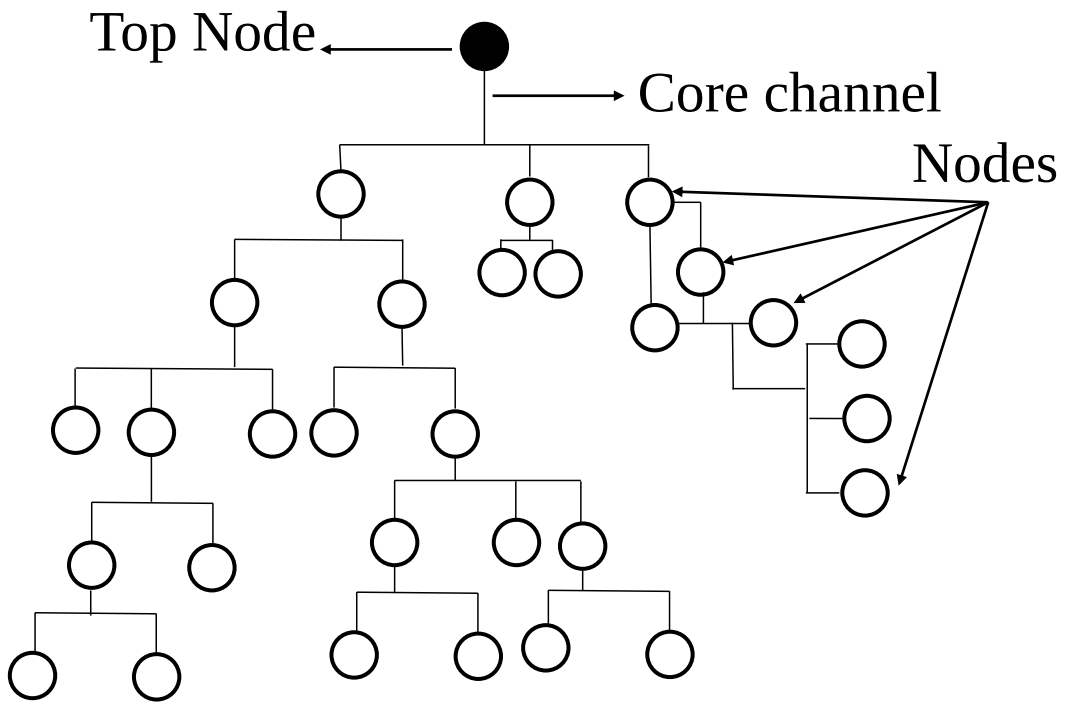

Source: Author´s own elaboration.

As figure 5 shows, classical political regimes impose themselves upon human social systems in a top-down fashion. They do so, in principle, in order to accomplish the function of the regime of organizing human social systems and to make easier the implementation of the authoritarian decisions that result from political systems computational processes.

**Figure 5. Political regimes imposing upon human social systems.**

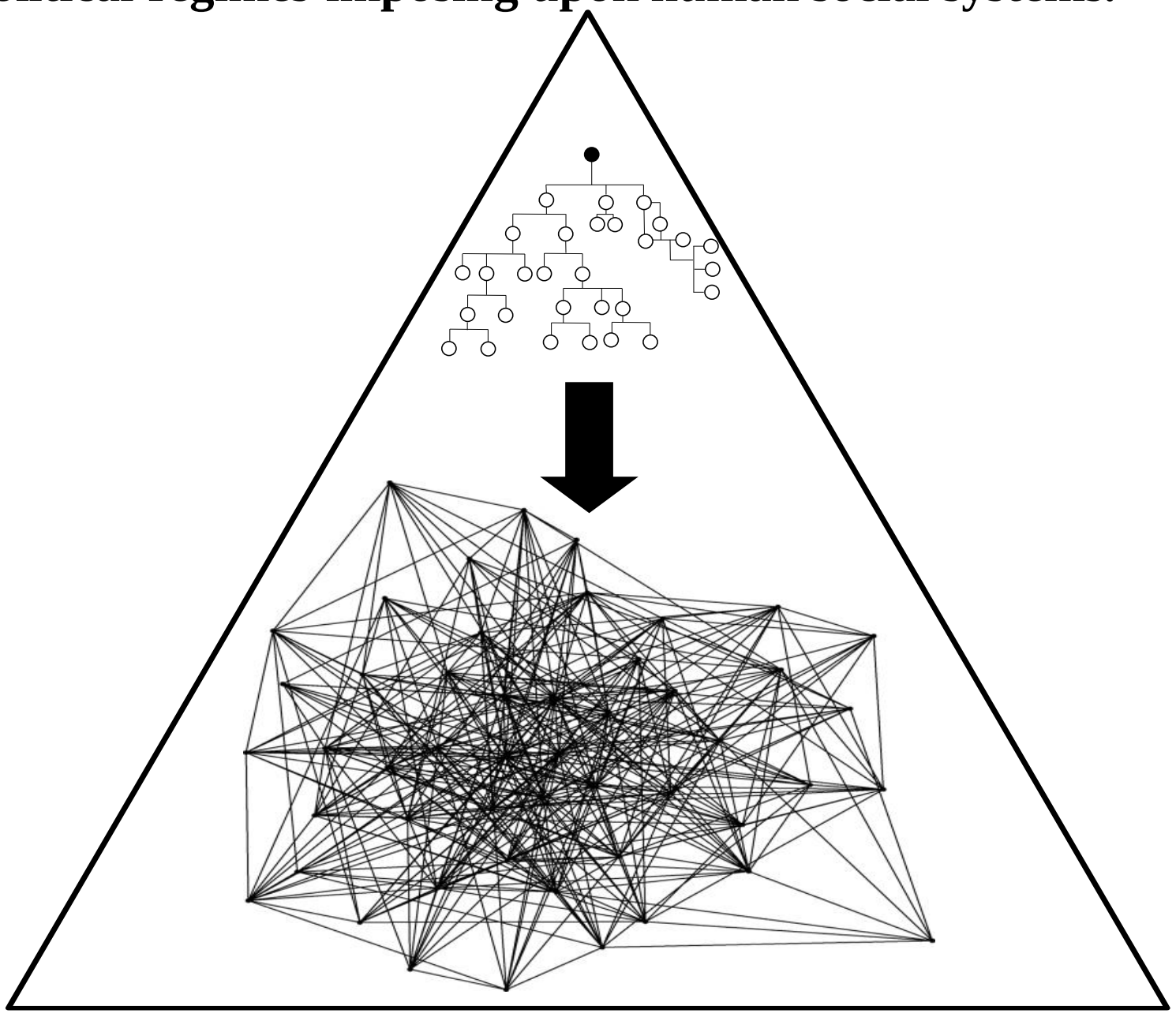

Source: Author´s own elaboration.

## 1.3. COMPLEXITY

Complexity is a scientific problem[17] that classical science set aside for most human history. Classical science seeks certainty, control, equilibrium, predictability and assumes direct cause-effect relations between inputs and outputs, whilst complexity makes itself possible thanks to uncertainty, unpredictability and non-linearity. Thereupon, complexity goes beyond the reductionism that characterizes classical science, which eliminates every possible noise or unwanted data from its deterministic models[18].

Complexity can be understood as a synonym and a result of nonlinearity. Nonlinearity implies non-proportioned (linear) cause-effect relations between the interactions within a system and their emergent patterns. Briefly, nonlinearity means that emergent results cannot be straightforwardly traced back to the properties and characteristics of separated entities that made them possible.

On the other side, the Sciences of Complexity are a group of sciences that study complex systems. Among them: Fractal Geometry, Chaos Theory, Non-Classical Logics,

[17] Compare Maldonado, Carlos Eduardo. "La Complejidad es un Problema, no una Cosmovisión". *UCM Revista de Investigación*. Vol. 13. (2009).
[18] Compare too Mitchell. *Complexity: A Guided Tour*.

Non-Equilibrium Thermodynamics and Artificial Life[19]. Complex systems are systems composed by many interconnected elements interacting in nonlinear fashions, meaning that interactions taking place in one part of the system may have nonlinear effects in the global structure or in other parts. Briefly, complex systems divide in two. Systems whose complexity maintains over time -like tornadoes, hurricanes and clouds- and systems with increasing complexity –or complex adaptive systems. The latter refer to living organisms and systems that present life-like properties, such as the immune system, cats, humans and human social systems.

Figure 6 shows the road that scientific knowledge has taken to recognize the existence of complexity. At first, classical science studied systems by analysis, i.e., by dividing them into their component elements and considering them as a sum of the latter. In the mid-twentieth century, *general systems theor*y and the discipline of cybernetics arose with principles that contrasted the analytical approach of classical science because while the latter understands systems in terms of their elements, general systems theory and cybernetics are more interested in holistic approaches.

The main problem with classical science, general systems theory and cybernetics is that the three ignore that when it comes to complex systems it is not possible to have complete knowledge about every state of them. And even less, on every moment of time[20]. The sciences of complexity overcame many of the shortcomings of its predecessors and have now become cutting-edge science.

Morin´s complex thinking[21] is closely related to the sciences of complexity. However, there are epistemological differences between both, like the computational apparatus that supports the sciences of complexity and the integrator aspect of complex thinking.

[19] Compare Maldonado, Carlos Eduardo & Gómez Cruz, Nelson Alfonso. *El Mundo de las Ciencias de la Complejidad.* 2011.
[20] Comprare Maldonado & Gómez. *El Mundo de las Ciencias de la Complejidad.* p. 51.
[21] Compare too Morin, Edgar. *El Método I, La Naturaleza de la Naturaleza.* 2006; Morin, Edgar. *El Método II, La Vida de la Vida*. 2006; Morin, Edgar. *El Método III, El Conocimiento del Conocimiento*. 2006; Morin, Edgar. *El Método IV, Las Ideas*. 2006; Morin, Edgar. *El Método V, La Humanidad de la Humanidad, La Identidad Humana*; Morin, Edgar. *El Metodo VI, Ética*. 2006.

**Figure 6. The road of the mainstream of science towards complexity.**

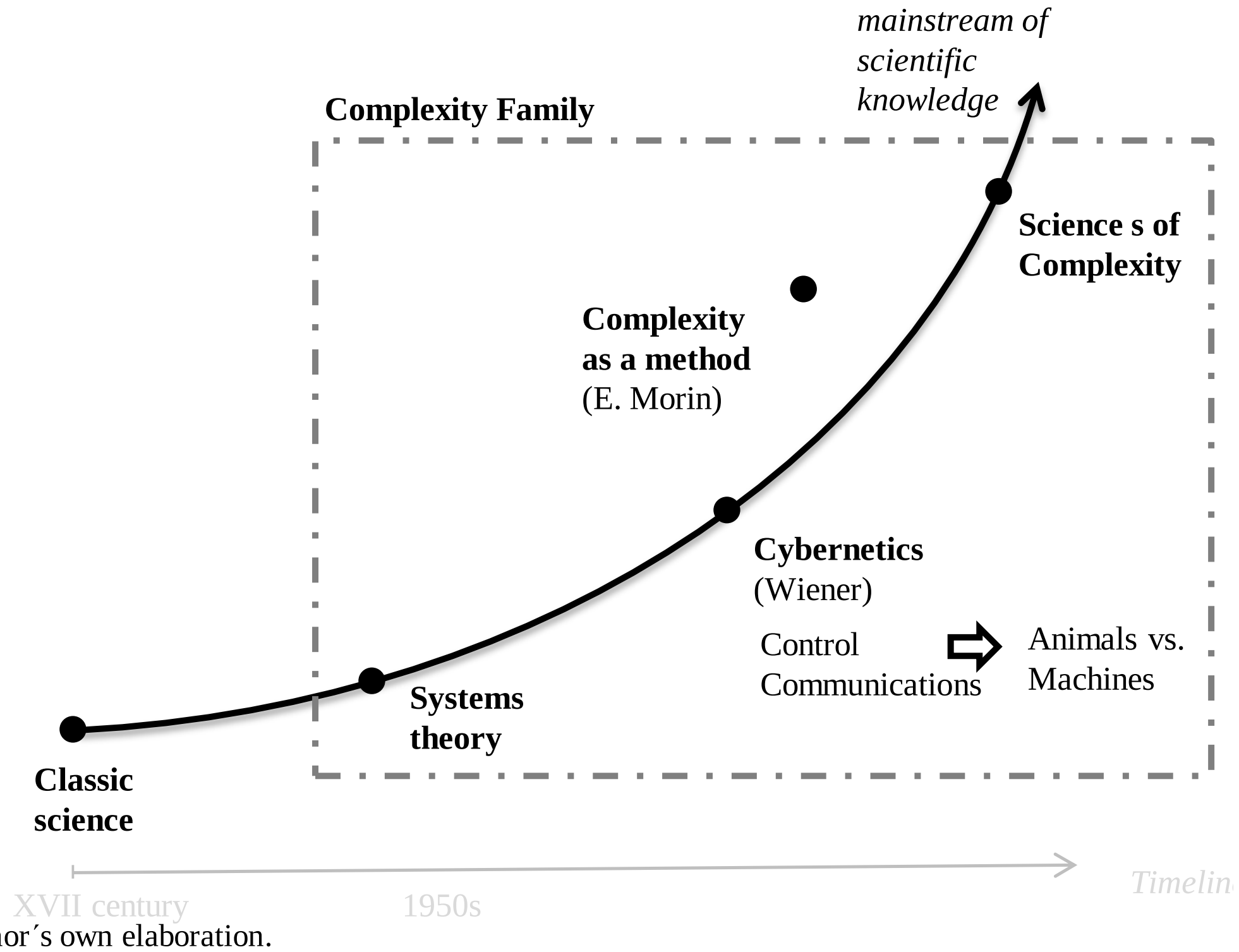

Source: Author´s own elaboration.

## 1.4. SELF-ORGANIZATION

Self-organization is the capacity that complex systems have for organizing themselves without any central or external control, giving rise to global patterns that emerge out of local interactions[22]. One example of self-organization is task allocation in harvester ants. These ants use chemical signals with their closest neighbors to decide if they should collect food, work in colony-related issues, patrol (or *hang around)*[23]. Contrary to the common misbelief, the queen ant does not control the colony. Her only task is to lay eggs. In fish schools, every fish follow what the neighbors do but there is not a fish that leads the collectiveness[24]. Table 1[25] shows examples of self-organized patterns.

[22] Compare Camanzine et.al. *Self-Organization in Biological Systems*; Haken, Hermann. *Information and Self-Organization: A Macroscopic Approach to Complex Systems.* 2006.
[23] Compare Gordon, Deborah. "Dynamics of Task Switching in Harvester Ants". *Animal Behavior*. Vol. 38. No.2.(1989).
[24] Compare Camanzine et.al. *Self-Organization in Biological Systems.*

**Table 1. Self-organization in natural and living systems.**

| | |
|---|---|
| 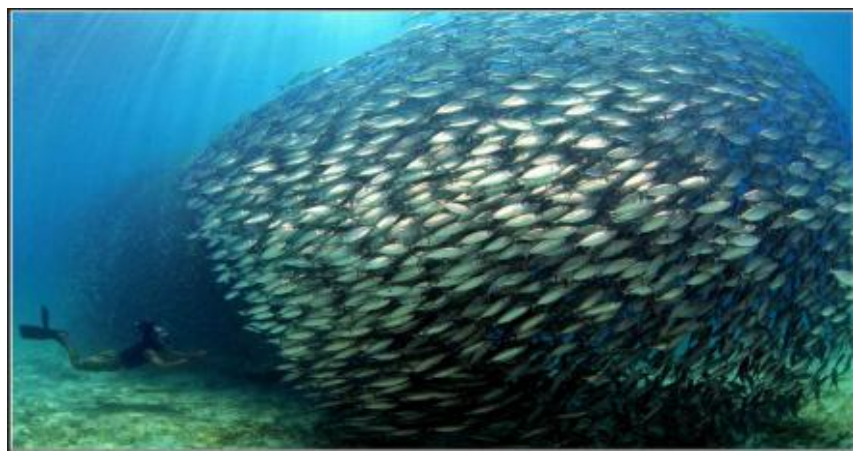 | 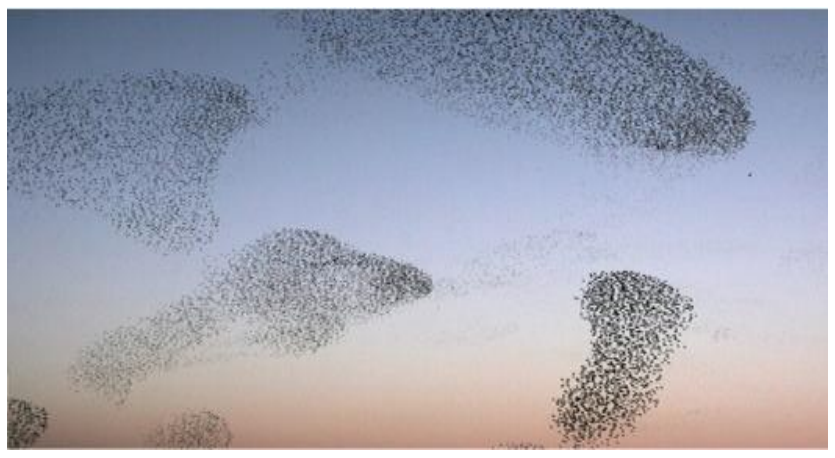 |
| 1.a. Fish School | 1.b. Flock of Starlings |
| 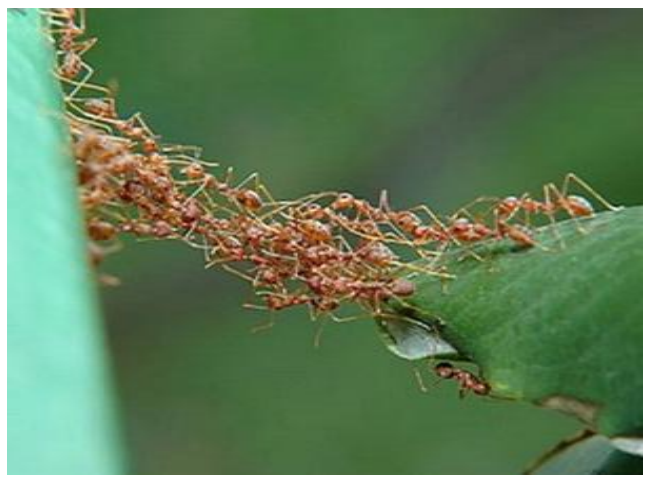 | 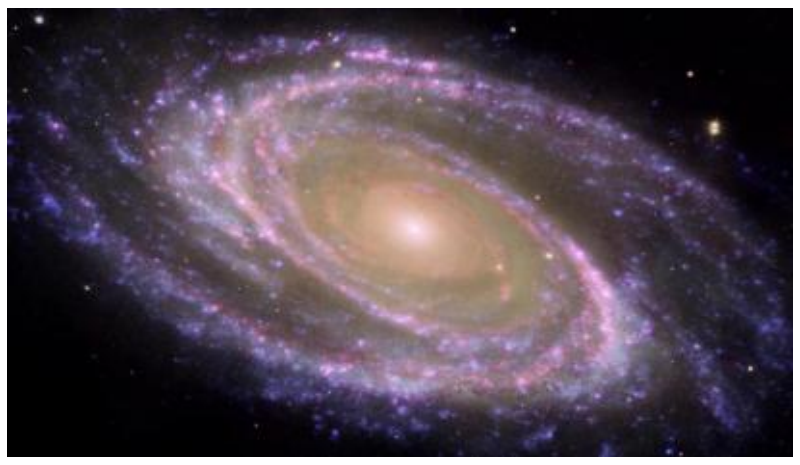 |
| 1.c. Ant Swarm | 1.d. Galaxy M81 |

Source: Author´s own elaboration.

A common example of self-organization in humans is friendship ties in social networks: community structure is an emergence of individual and local interactions pursuant to particular interests or needs.

On the other hand, sociopolitical self-organization is the name given in this monograph to the political dynamics in human social systems that emerge autonomously in interactions that develop using local information and do not obey to any general ruling principle –or government. It is claimed here that sociopolitical self-organization will lead to the emergence of organized global patterns that evolve in harmonic, but dynamic, flows in human social systems; and that this is the best way for organizing them.

## 1.5. COMPLEX NETWORKS

"A network (graph) is a diagrammatic representation of a system".[26] Networks are conformed by elements (nodes) and their connections (links or interactions) giving rise to particular relations, in an arrangement that is called topology[27]. The topological –or structural- properties of networks started to be studied in mathematics, in the field of Graph theory. Nonetheless, this monograph does not discuss the mathematical aspect of them. Instead, it focuses on the distribution of the nodes in networks and the kind of power relations that arise from the interactions between the nodes. Precisely, how tree topologies in political regimes imply command relations for the organization of human social systems and how sociopolitical self-organization can give rise to global complex network patterns.

[25] "Fish school". Electronic document; "Flock of starlings". Electronic document; "Ant swarm". Electronic document; "Galaxy M81". Electronic document.
[26] See Estrada, Ernesto. *The Structure of Complex Networks.* 2012. p.4.

Complex networks (figure 7) [28] are scale free networks, which means that the connectivity between their nodes is feasible to be described by a power law (figure 8). Power-laws in complex networks imply that there are few nodes, called hubs, with more connections than average nodes. For instance, in networks of sexual interactions among humans, there are less people who have had many sexual partners than those who have had average or few. Thus, contagious disease spreading patterns have this type of structure[29]. The topology of Internet can be described using a power-law too: there are few highly connected webpages in comparison to those that are just subtly connected. Figure 9, *a picture* of Internet webpages, illustrates this idea.

**Figure 7. Complex network.**

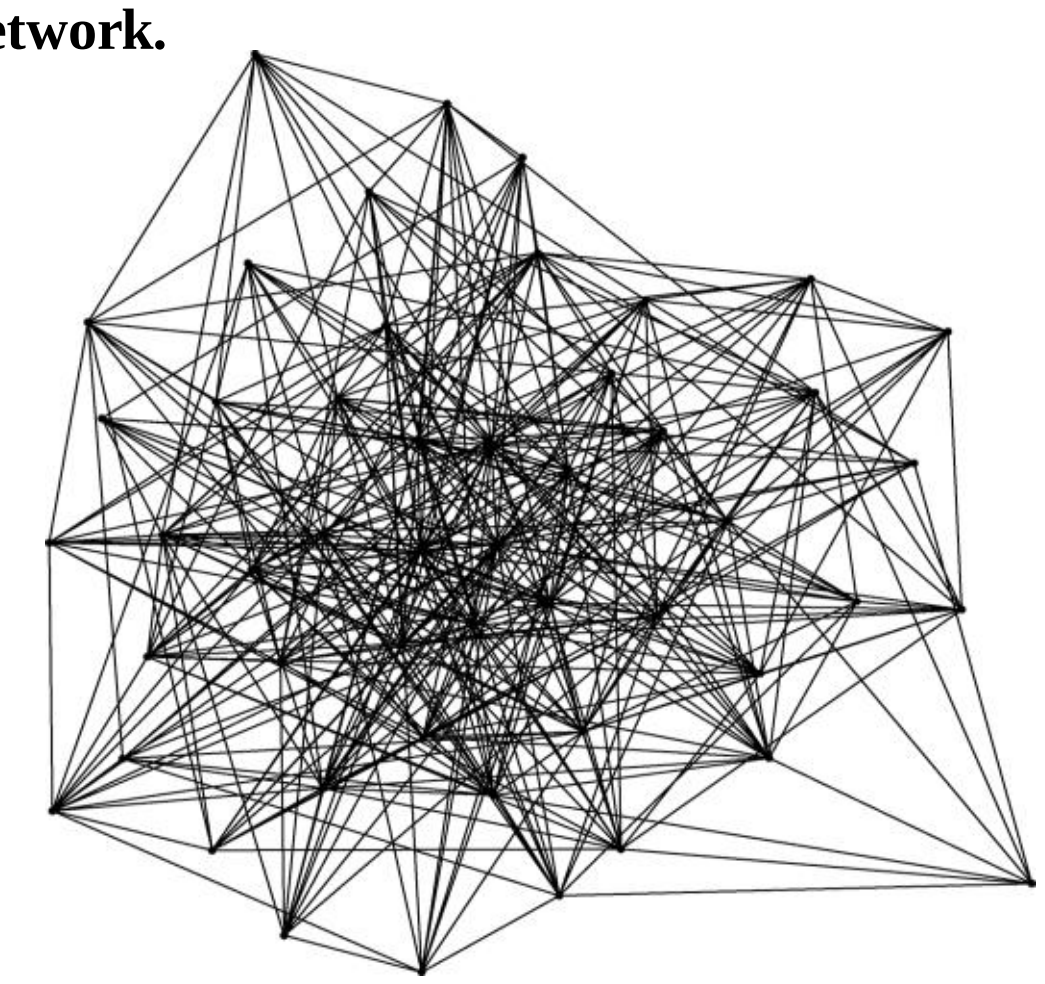

Source: Prettejohn, Berryman and McDonnell. In Methods for Generating Complex Networks with Selected Structural Properties for Simulations: a review and tutorial for neuroscientists. 2011.

[27] Compare Oppenhimer, Priscila. *Top-Down Network Design.* 2004.

[28] See Prettejohn, Brenton et.al. "Methods for Generating Complex Networks with Selected Structural Properties for Simulations". *Frontiers in Computational Neuroscience.* Vol 5, No. 11 (2011).

[29] Compare Jones, James Holland and Handhock, Mark. "An Assessment of Peferential Attachment as a Mechanism for Human Sexual Network Formation". *Proceedings of the Royal Society of Biological Sciences.* 2003.

Complex networks started to be widely studied one decade ago, after understanding that there was a particular network structure present in many information processing dynamics of living systems and in systems that present life-like properties when they self-organize. For instance, in networks of protein interaction[30], food webs[31], Internet´s structure[32], networks of online music recommendation[33] and in social networks[34], as well.

**Figure 8. Power-Law.**

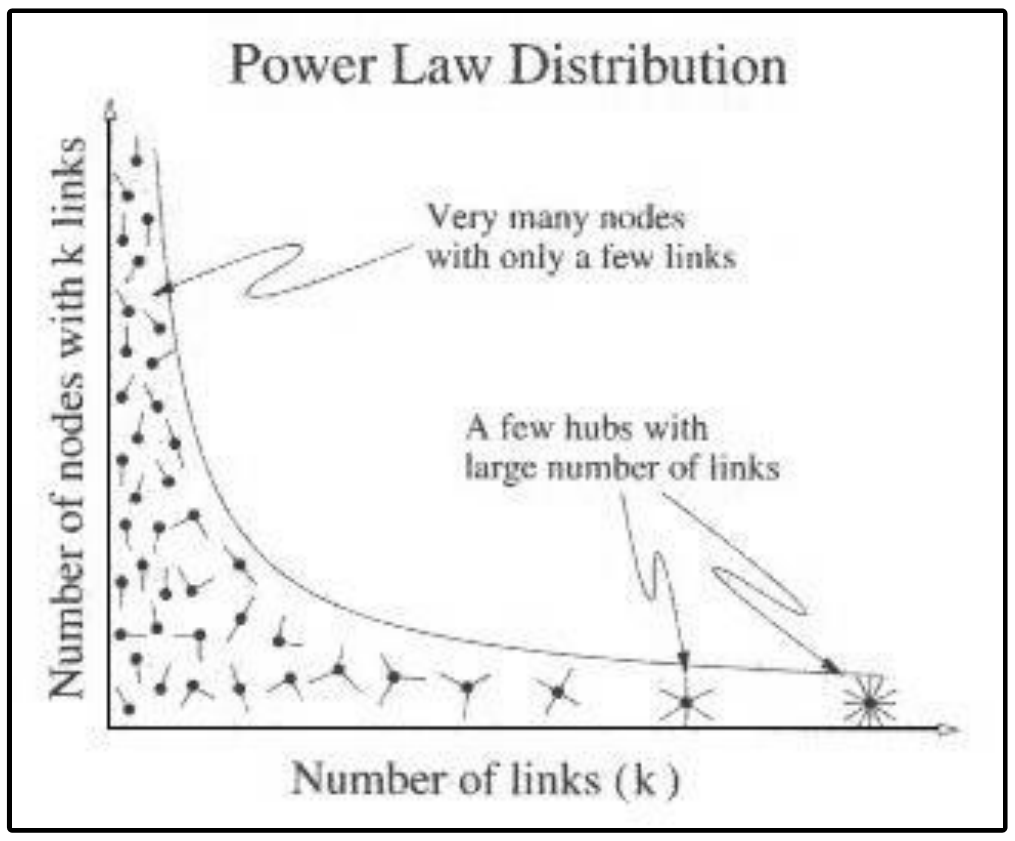

Source: Electronic document.

**Figure 6. Hubs in Internet.**

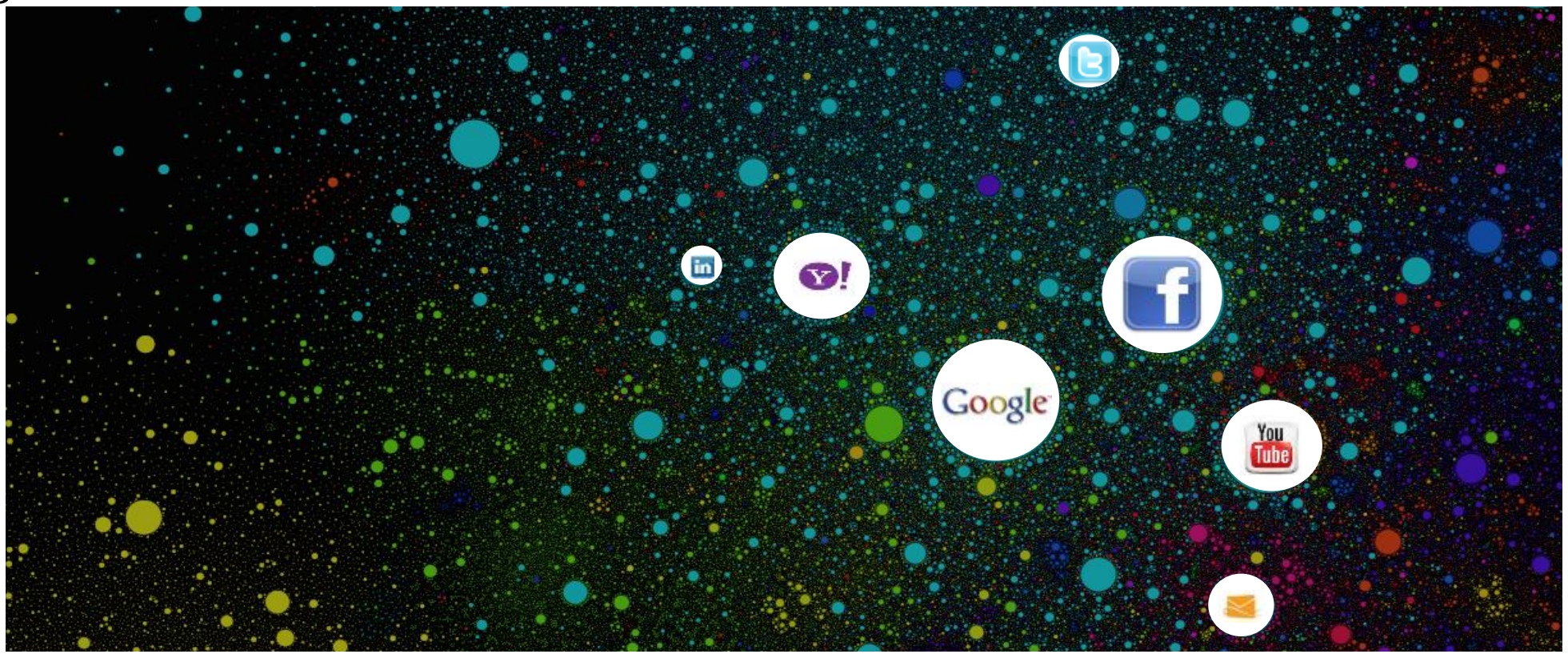

Source: Electronic document.

[30] Compare too Kuchaiev, Oleskki et.al. "Topological Network Alignment Uncovers Biological Function and Phylogeny". *Interface.* Vol. 1 No. 4. (2010).
[31] Compare too Strogratz, Steven. "Exploring Complex Networks". *Nature, Vol.* 410. (2011)
[32] Compare too Estrada. *The Structure of Complex Networks.*
[33] Compare too Cano, Pedro et.al. "Topology of Music Recomendation Networks". *Chaos.* Vol. 16 (2006)
[34] Compare too Watts, Duncan. *Six Degrees. The Science of a Connected Age.* 2003.

Complex networks are made out of vastly diverse networks of networks and interactions (links or connections), that can be directed or not directed. A non-directed link means that if there are two nodes connected to each other, A and B, A influences B as much as B influences A. However, if the link is directed from B to A, only B would influence A. A real-life example of directed links is the basic structure of social networks, online or not. People tend to be friend with their friend´s friends. So, given three friends, Alice, Bob, and Martha (figure 10), they would form a non-directed triangle, since Alice is friend with Bob and Martha; Bob is friend with Alice and Martha; and Martha is friend with Alice and Bob. In respect to the direction of connections in networks, tree topologies, the structures of political regimes, are directed in a top-down fashion. Given that complex networks are the most natural nodal arrangement for interactions in human social systems when they self-organize, in this monograph it is claimed that the self-organization of sociopolitical interactions in human social systems would make complex network patterns to emerge.

**Figure 10. Triangles in social networks..**

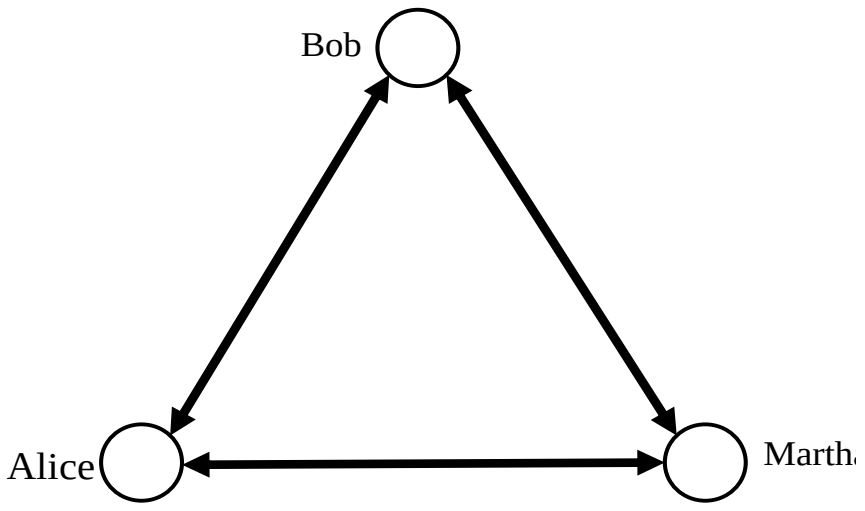

Source: Author´s own elaboration.

## 2. LITERATURE REVISION: POLITICAL REGIMES, STRUCTURAL PROPERTIES AND NETWORK APPROACHES

This chapter presents selected works found in a search conducted in the course of the year 2012, where the most relevant and scientifically influencing political science and complexity journals were explored. Mainly, the search sought to finding authors that discussed the structural and dynamic properties of political regimes in the contexts of the sciences of complexity, in relation to their function of organizing human social systems (in a top-down fashion), but that, at the same time, also addressed to the self-organization of human social systems –or their sociopolitical interactions-, considering the existence of political regimes, political systems and their top-down control nature. In respect to this quest, with few exceptions, very little has been said. Indeed, no authors were found that directly discussed the subject. Therefore, this literature revision will emphasize on those authors that have studied structural properties of organizations and that, from a general perspective, could enlighten the problem of how political regimes unsuccessfully try to organize human social systems by imposing  that impose top-down control upon them.

Most of the bibliography related to the architecture of political regimes was either (a) interested in the nature of the elements in the network (i.e., their position in government and how they get there); or (b) those centered on the types of relations between the elements (subordination, command, etc.). However, they did not emphasize on their influence over human social systems or how structure and dynamics are mutually influenced by each other. From a Political Science perspective, the name given to (a) and (b) were dispositional and relational properties[35], respectively; classical network theory named them structural and logical topology[36]; whereas the study of complex networks describe them as structural and dynamic properties[37].

---

[35] Compare Elgie, Robert. "The Classsification of Democratic Regime Types: Conceptual Ambiguity and Contestable Assumptions" *European Journal of Political Research*, 1998.
[36] Compare Oppenheimer. *Top-Down Network Design.*
[37] Comprare Aldana, Maximino. *Redes Complejas.* 2006. Electronic document.

In respect to these properties, *Robert Elgie*[38] argues that it would be preferable to study only one at the time. However, *Erik-Has Klijn*[39] claims that descriptions strictly focused on relational properties are incomplete because they do not differentiate between the system and its structure. *Albert-Lazló Barabasi*[40], one of the precursors of network science, states that the study of networks would not be complete if it only focuses on structural properties without deepening into how they influence the dynamics of the system. Following his statement, this monograph will use a combination (a) and (b), meaning (c), to study the implications of the topologies of political regimes in respect to their function of organizing human social systems.

A common place was to find authors for which transformations in the way in which humans are organized are equivalent to changes within electoral systems, leaders, *typology* of the regime (presidential, parliamentarian, proportional, etc.), among others[41], but not exactly any profound change.

*Örjan Bodin* and *Jon Norberg*[42] two of the few authors that studied the structure of regimes within a topological framework, in the sense of graph and network theory, centered their attention on information networks in organizations. However, they put too much attention over managerial units, which makes their perspective in favor of top-down control to distance from the one in favor of self-organization presented in this monograph. *Raymund Werle*[43] also studied the role of information in the structure of political regimes, making emphasis on how information networks –especially those facilitated by Internet- are extending national infrastructures.

---

[38] Compare Elgie, Robert. "The Classsification of Democratic Regime Types: Conceptual Ambiguity and Contestable Assumptions" *European Journal of Political Research,* (1998). p.235.

[39] Compare Klijn, Erik-Has. "Analyzing and Managing Policy Processes in Complex Networks: A Theoretical Examination of the Concept Policy Network and its Problems". *Administration & Society* (1996).

[40] Compare Barabasi, Albert-Lazló. *Linked: How Everything Is Connected to Everything Else and What It Means for Business, Science and Everyday Life.* 2003.

[41] Compare Colomer, Joseph M. "Desequilibrium Institutions and Pluralist Democracy". *Journal of Theoretical Politics.* Vol 13. No.3 (2001)

[42] Compare Bodin, Örjan and Jon Norberg. "Information Network Topologies for Enhanced Local Adaptive Management". *Environmental Management*. Vol.35, No. 2 (2005).

[43] Compare Werle, Raymund. "The Impact of Information Networks on the Structure of Political Systems". In *Understanding the Impact of Global Networks on Local Social, Political and Cultural Values*. 1999.

*Göktuğ Morçöl* and *Aroon Wachhaus*[44] claim that the Structuralist Theory of Anthony Giddens can provide a link between the sciences of complexity and political regimes, (in general, public administration), since it uses notions of unpredictability and stability, at the same time. On the contrary, David Knoke[45] suggests that network models in public administration enriched by graph topology allow studying better Gidden´s social structures. Although Giddens did not have topology in his theory of Structuralism, Knoke´s idea could reinforce the problem addressed here: how the tree topologies of political regimes impose top-down control over human social systems for organizing them, hampering the self-organization of human sociopolitical interactions.

Regarding how the structural properties of political regimes influence their function of organizing human social systems (c), there are three important concepts that should be discussed: hierarchy, heterarchy and anarchy. Hierarchies are just one of the many characteristics that the structures of political regimes present. Hierarchical regimes usually have a supreme ruler in the top of the structure (one individual or a group) that directs the flow of information in the system. According to *Gerard Fairtlough*[46], for traditional political thinkers, such as Thomas Hobbes or Max Webber, the absence of hierarchies is inconceivable because it is though that it would give rise to one of the most feared concepts in political science: anarchy. This may be why many political scientists favor physical coercion coming from political regimes: in order to guaranty not to go back to that theoretical stadium of absence of order, where individuals were not ascribed to any State or ruled by any coercive political authority. It is also why many accept self-organization as the price to pay for having the certainties over the future that the linear dynamics of tree topologies provide. Despite this common misbelief, it has never been proven that hierarchies produce more order in complex environments. In fact, if this were to be true, human social systems -historically characterized by being ruled hierarchically- would be completely organized and there would not exist any agent acting outside the borders of the establishment or against it. No guerillas, paramilitary institutions, rebellion and not even strikes[47]. It is also widely thought that hierarchies enable leadership to emerge, which would be valid if *to emerge* meant *to be imposed* or *to be elected.* And neither imposition nor election are ways of emergence, since none of them imply synergetic processes based on self-organized interactions.

[44] Compare Morçöl, Göktuğ & Wachhaus, Aroon. "Network and Complexity Theories: A Comparison and Prospects for a Synthesis". *Administrative Theory & Practice.* Vol. 31, No. 1 (2009).
[45] Compare Knoke, David. *Political Networks: The Structural Perspective.* 1990.
[46] Compare Fairtlough, Gerard. *The Three Ways of Getting Things Done.* 2006.

*Carole Crumley*[48] divides hierarchies into scalar and control. In a scalar hierarchy every level can be affected by the others. This is the type of hierarchy present in complex systems. A biological organism, for instance, can die if there is a failure in one of its subnetworks or internal structures, such as organs or groups of cells. However, one organ can stop functioning if the whole organism undergoes an extreme situation of physical stress, such as hypothermia. Similarly, one small group of individuals interacting online can trigger large offline sociopolitical revolutions but, at the same time, the latter can stimulate that online groups join or support them. On the other hand, control hierarchies, the hierarchies of non-complex systems, are those hierarchies where top-levels are not affected by lower-levels. Political regimes have control hierarchies, since their structure is maintained by means of top-down and authoritarian mechanisms.

*Michael North* and *Charles Macal*[49] divide hierarchies into noisy hierarchies and pure hierarchies. Noisy hierarchies are characterized by presenting noise or random errors in the transmission of information among levels, whereas pure hierarchies have no loss of information during the transmission from the top to the bottom. As the authors affirm, complex systems present noise, errors and uncertainties. Therefore, there is always noise and uncertainty in the computational dynamics of decision-making processes. Assuming linear relations in the transmission of information among hierarchies in political regimes when they try to organize human social systems is not asserted. This is one reason why self-organization is the best alternative for organizing human social systems: there is no need to coordinate huge amount of information that brings more noise into decision-making processes.

---

47 Compare Mezza-Garcia. Bio-Inspired Political Systems.

48 Compare Crumley, Carole L. “Heterarchy and the Analysis of Complex Societies”. *Archeological Papers of the American Anthropological Association. Special Issue: Heterarchy and the Analysis of Complex Societies.* Vol. 6, No. 1 (1995).

49 Compare North, Michael, y Charles M. Macal. *Managing Business Complexity.* 2007.

The term *heterarchy* was extrapolated to human and social sciences from its original meaning in neuroscience, with a work by *Warren S. McCulloh*[50]. Heterarchical organizations have network structures, not pyramidal forms, and distribute the control horizontally within various actors -this decentralized organization is shared among complex systems. *Edward O. Wilson* and *Burt Hölldobler* refer to the organization of ant colonies as heterarchies because ants are all connected to each other, meaning that there is not a hierarchical organization[51]. In a heterarchical organization there are dynamic and distributed processes of synergetic decision-making. Hence, one of its advantages is the possibility for creative solutions.

Anarchy, the third concept, has been recently used to characterize relations among political regimes. In new public management it has been a metaphor for theorizing about how public organizational networks should not be central-based for better policy-making[52]. There are, indeed, strong solid bonds between anarchy and network governance. Networks of policy-making were proposed as an alternative to traditional models of hierarchical control. Networks in new public management refer to interdependence arrangements that do not follow top-down patterns of nodal connections[53] or hierarchical positioning of them[54]. This discussion of networks is not based, however, on the field of complex networks. One historical reason for this is that the theory of complex networks was developed decades after the concept of networks in public administration became popular. In accordance with Franz Pappi and Christian Henning "policy networks are often used as a metaphor to describe new forms of governance beyond state control involving both public and private actors".[55] Nonetheless, this networks are still far of having implicit the notion of self-organization in human social systems because they are still based on traditional tree topologies, which operate with top-down control. Naim Kapuku´s figure[56] (figure 11) shows this relation.

[50] Compare McCulloh, Warren, S. "A Heterarchy of Values Determined by the Topology of Nervous Nets". *Bull. Math. Biophysics.* Vol. 7 (1945).
[51] Compare Wilson, Edward O. and Hölldobler, Burt. "Dense Heterarchies and Mass Communication as the Basis of Organization in Ant Colonies". *Ecology and Evolutions.* Vol.3, No.3 (1998).
[52] Compare Wachhaus, Aroon. "Anarchy as a Model for Network Governance". *Public Administration Review.* Vol.72, No. 1 (2011).
[53] Compare O´Toole, Laurence and Meier, Kenneth. "Desperately Seeking Selznick: Cooptation and the Dark Side of Public Management in Networks". *Public Administration Review.* Vol. 64, No. 6 (2004).
[54] Compare O´ Toole, Laurence. "Shaping Formal Networks Through the Regulatory Processes". *Administration and Society.* Vol. 36, No. 2 (2004).
[55] See Pappi, Franz Urban and Henning, Christian. "Policy Networks: More than a Metaphor?". *Journal of Theoretical Politics.* Vol. 10 (1998).

**Figure 11. Interorganizational networks.**

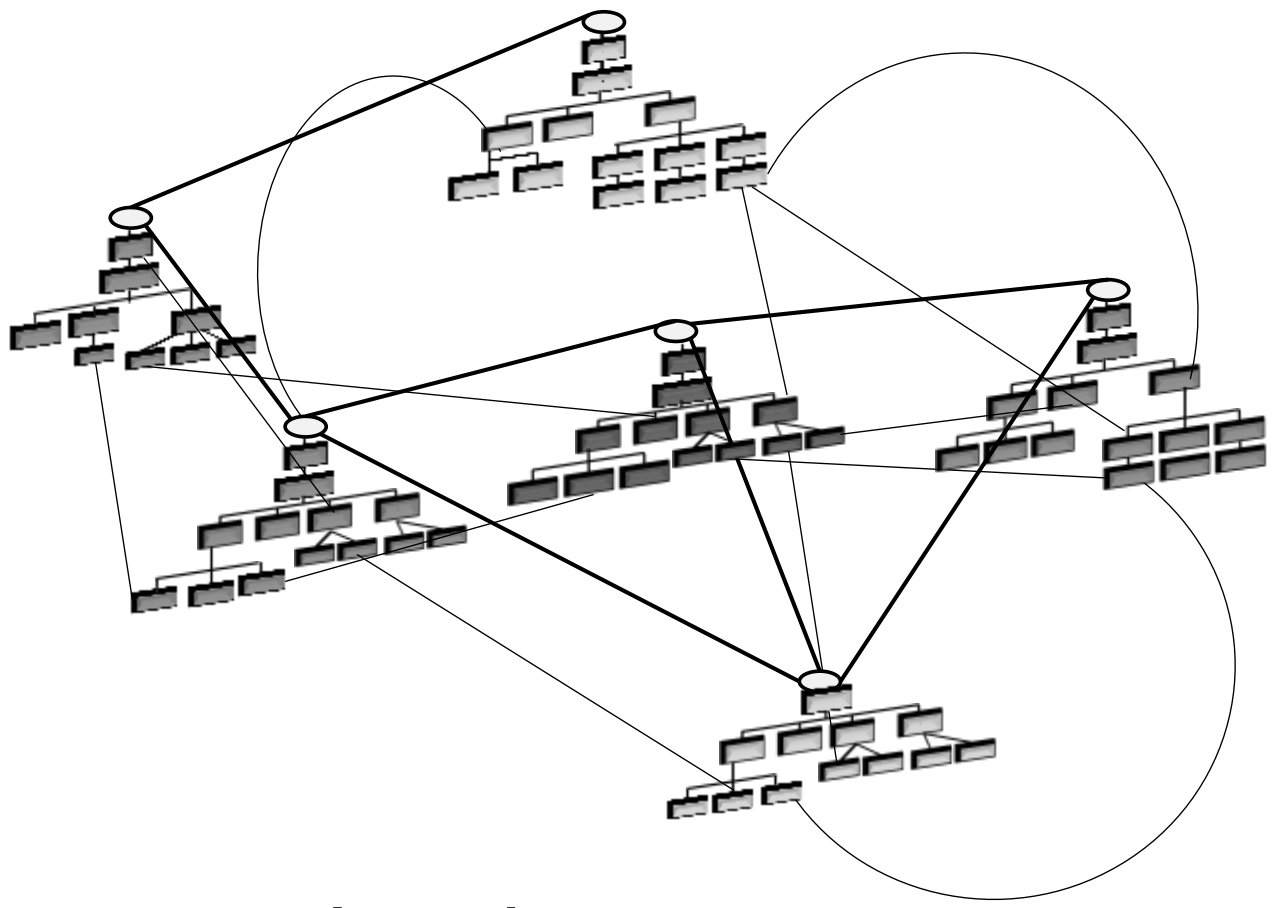

Source: Naim Kapuku. Interorganizational networks. In Interagency Communication Networks During Emergencies. 2011. p.211

The problem with the dynamics that figure 11 entail is that although the connections between the tree topologies might be self-organized, tree topologies are ruled by the principles of classical network theory, where the result of connecting tree topologies is, simply, tree topologies connected together, thus, the emergent global pattern is not self-organized. *Aroon Wachhaus*[57] states that the applications of network theory in policy should transcend in a much deeper fashion to governments because most of the network language used today in political theory remains inside the scope of top-down control, where networks are the echo of hierarchies. Thereby, Wachhaus proposes an interplay between networks and anarchy for increasing the possibilities of emergent cooperation networks. Wachhaus´ perspective is completely valid. After all, as it will be pointed out in the next chapters, anarchy bases on the idea of self-organized interactions.

According to *Stephen Goldsmith* and *William Eggers*[58], networks in public management arose in times of increasing complexity in societies, when complexity overwhelmed hierarchical models. Meek et al. argue that one possible reason for this is that

[56] See Kapucu, Naim. "Interagency Communication Networks During Emergencies". *American Review of Public Administration.* Vol.36, No. 2. (2006). p.211.
[57] Compare Wachhaus, Aroon. "Anarchy as a Model for Network Governance". *Public Administration Review.* Vol.72, No. 1 (2011).
[58] See Goldsmith, Stephen, and Eggers, William. *Governing by Network.* 2004. p.7.

“rigidly hierarchical organizations directed trough top-down decision-making is likely to be ineffective”[59]. Similarly, *Peter Bogason* and *Julieth Musso*[60] state that hierarchies, even if they are decentralized, cannot handle the complexity of contemporary world. Bogason and Musso´s argument serve as a foundation for letting human social systems to self-organize, because self-organization does not require that the elements that interact –in this case, individuals and sociopolitical groups- have global information or coordination. *O´Toole* and *Meier*[61] claim that this complexity forces policy networks to add more actors to regulatory processes, tilting the balance of power, and making those regulatory processes to complexify[62]. Not in a pejorative sense, though. For this reason, it could be said that public management is turning to more anarchic forms of policy making, since having one central controller is starting to be conceived as incompatible with the complexity of human social systems.

Some management authors, such as *Eve Mitleton*[63] and *Robert Lewin et al*.[64] recognize that organizations have nonlinear interactions carried out by numerous elements that are continuously changing and adapting, thus, they state that organizations are complex adaptive systems. If political regimes were conceived like this, then the possibility for the self-organization of human sociopolitical interactions would easily arise because complex adaptive systems are continuously exchanging information with their environments. This could imply that the traditional boundary that exists between the political society and the civil society would blur.

An important theme of discussion for the literature revision is the study of the role of information networks in the structures of political regimes. The reason is that in this monograph it is claimed that political systems are computational systems due to how they process information when transforming inputs into decisions in order to achieve their function of organizing human social systems. *Christian Fuchs*[65] was, by far, who came closest to the problem addressed in this work, despite that he centers his attention in political systems and not in their relation to human social systems. Fuchs argues that political systems are formed by certain structures that organize power by permitting and constraining the behavior of individuals in human social systems. He also states that thanks to the complexity that characterizes political systems, they are self-organized in the form of arrangements –i.e., structures and rules- that dynamically develop, change and adapt. However, Fuchs gives too much credit to the current role of self-organization in political systems –which, in reality, is not as self-organized as the paper exposes, because great part of the dynamics of political systems obey to top-down arrangements, instead of bottom-up synthesis. *Peter Dittrich* and *Lars Winter*[66] went a step farther designing a chemical-based catalytic network model for understanding how political systems process information and studied possible hidden structures in the institutions in charge of the decision-making processes in political systems, which lead them to assimilate political systems with a chemical reaction. This approach is very similar to a paper by *Gary Gemmill* and *Charles Smith*[67], who proposed a model that supports the need of letting human sociopolitical interactions self-organize in their dynamic environments. They stated that organizations are not static, thus, could be assimilated with Ilya Prigogine´s dissipative structures[68]. That is, complex structures that exist in non-equilibrium environments and are continuously exchanging matter and energy (information) with them, acquiring their dynamic stability from that process. Gemmill and Smith show interest in how transformations occur in organizations and although they do not refer to any specific type of institution, their emphasis on how organizations react as dissipative structures when facing structural transformations after suffering perturbations coming from the outside is very enlightening. After all, as Camanzine et al claim, “open systems, in which there is a continual influx of energy or matter, reactions occur far from chemical equilibrium, and structures emerge through interactions obeying nonlinear kinetics”.[69]

---

[59] See Meek et.al. “Complex Systems, Governance and Policy”. *Emergence: Complexity and Organizations.* Vol.1, No. 2 (2007). p.25.

[60] See Bogason, Peter and Musso, Julieth. “The Democratic Prospects of Network Governance”. *The American Review of Public Administration.* Vol. 36, No. 3 (2006). p.14.

[61] Compare O´Toole, Laurence and Meier, Kenneth. “Desperately Seeking Selznick: Cooptation and the Dark Side of Public Management in Networks”. *Public Administration Review.* Vol. 64, No. 6 (2004).

[62] Compare O´ Toole, Laurence. “Shaping Formal Networks Through the Regulatory Processes”. *Administration and Society.* Vol. 36, No. 2 (2004).

[63] Compare Mitleton-Kelly, Eve. *Complex Systems and Evolutionary Perspectives of Organizations: The Application of Complexity Theory to Organizations.* 2003.

[64] Compare Lewin et.al. “Complexity Theory and the Organization: Beyond the Metaphor”. *Complexity.* Vol.3, No. 4 (1998).

---

[65] Compare Fuchs, Christian. *The Political System as a Self-Organizing Information System.* In *Cybernetics and Systems.* 2004.
[66] Compare Dittrich, Peter, y Lars Winter. “Chemical Organization in A Toy Model of the Political System” ECCS´07. 2007.
[67] Compare Gemmill, Gary, and Smith, Charles. “A Dissipative Structure Model of Organization Transformation”. *Human Relations.* Vol. 38, No. 8 (1985).
[68] Compare too Prigogine, Ilya. *Thermodinamic of Irreversible Processes.* 1955.

For closing this chapter, it is important to stress that, on one side, complexologists have studied organizations but barely within a political framework. On the other side, it is clear that scholars of politics and complexity continue to study political regimes without deeply considering how their structural properties affect the dynamics and the tendency to self-organize of human social systems. Furthermore, some theorist find limitations in the traditional organizational structures of political regimes, but most of them are still thinking in organizing social systems by means of top-down control. In sum, there is a conceptual vacuum here given that the function of political systems and their regimes is finding optimal solutions to the organization of human socials systems, but political science has not been very concerned about looking for better ways to organize the latter, despite the many disadvantages of top-down control -which will be listed in the following chapter. In addition to this, political science has not approached very profoundly the sciences of complexity for acknowledging the complex nature of the systems that political regimes organize. It is not surprising to find sparse literature that properly addresses how political regimes (and political systems) do not allow the self-organization of human social systems, and that when it comes to visualizing the complexity of their sociopolitical interactions, the majority of authors ignore that such complexity arises from the complex nature of the individuals and their interactions, but not precisely because political regimes know how to deal with complexity. This is why most political science authors are still thinking in terms of governments, public policies, laws, election and voting dynamics.

[69] See Camanzine et.al. Self-Organization in Biological Systems. p. 29

# 3. COMPLEXITY IN HUMAN SOCIAL SYSTEMS

This chapter presents the necessity of opening political regimes to the self-organization of human social systems. It is claimed that political regimes should be complexified until reaching a point where there is not a difference between the political and the civil society. In the short term, this could be seen as gradualist. However, it should be understood within a *longue durée* proposal, intended to avoid disruptive situations that might come after amplified sociopolitical fluctuations. The second section of the chapter discusses sociopolitical self-organization starting from the theory of self-organization in biological systems.

## 3.1. EXTRAPOLATING THE LIFE-LIKE PROPERTIES OF HUMAN SOCIAL SYSTEMS TO POLITICAL REGIMES: THE RISE OF SOCIOPOLITICAL SELF-ORGANIZATION

Human social systems are complex systems characterized by nonlinear and diverse interactions. They can be described in terms of biological systems because they are systems that present life-like properties. For instance, they are open, dynamic, adaptive, far-from-equilibrium and present evolutionary dynamics, thus, they can learn, are continuously changing and by no means their future states are previously fixed. Indeed, their decentralized nature makes them close to being anarchic entities because they can make organized patterns to emerge without any top-down control or central authority. Additionally, they are capable of synthetizing their dynamics of self-*control intrinsically to their self-organization.

In contrast with the life-like properties of human social systems, classical political regimes do not self-organize (besides not letting human social systems do so) because they are linear, instead of non-linear; they cannot be described in terms of biological properties but using terms from classical physics (Newtonian physics[70]); they do not adapt; are very static; closed; and have defined boundaries with their environments (human social systems). Additionally, rather than being far-from-equilibrium, they are stationary; centralized; cannot be described with life-like properties; and are teleologically oriented. But, most importantly, they do not self-organize. Coercion is the replacement of cooperation in political regimes; and their evolutionary dynamics are present in the form of lengthy and out of date formal processes –that are strictly hierarchical. This means that the organization of human social systems is made by systems that oppose their own dynamics. As a consequence, great part of the complexity that emerges from bottom-up interactions in human social systems is eliminated or, at least, restricted and constrained by them.

[70] Compare Ma, Shun-Yun. "Political Science at the Edge of Chaos? The Paradigmatic Implications of Historical Institutionalism" . *International Political Science Review.* Vol. 28, No. 1 (2007).

Clearly, when political regimes intervene in the organization of human social systems, they leave no space for self-organized sociopolitical dynamics to emerge –or for complexity to peacefully bifurcate. Given that the natures of human social systems and political regimes are too different, regimes push human social systems beyond the edge of chaos, where organization is just an unachievable goal (figure 12).

**Figure 12. The Edge of Chaos**

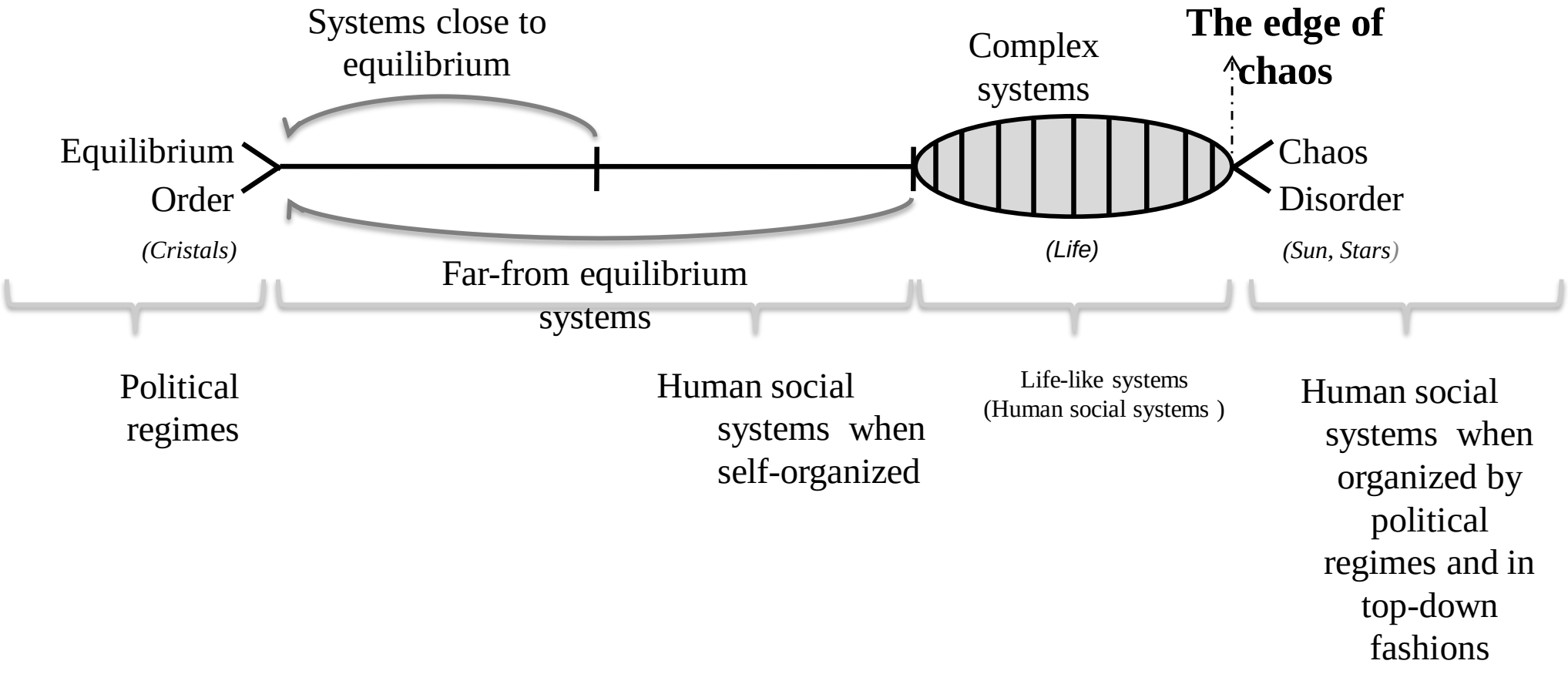

Source: Adapted from Maldonado, Carlos Eduardo and Gómez Cruz, Nelson Alfonso, *El Mundo de las Ciencias de la Complejidad.* (2011). p.15

For counteracting this situation, the complexity of political regimes could be augmented, for generating organized global patterns in human social systems. It should be clarified that the structure referred to here as emergent coherent *global* patterns is not related to any historically or power-related imposed idea such as nation-states (however, it could be a community) and that *local* interactions are not necessarily geographically-based. Coherent patterns can only be achieved if human social systems are allowed to self-organize, i.e., by means of the self-organization of their sociopolitical interactions. Because despite that self-organization occurs at the edge of chaos, the order that it produces is in the realm of the organized[71].

Table 2[72] presents some of the modes of complexity that could be considered for complexifying regimes. An interplay between these modes of complexity, augmenting the importance of citizen´s direct participation mechanisms and breaking the barrier there is between the political and the civil society could lead towards human social systems finally able to self-organize.

As it was suggested when mentioning the modes of complexity, increasing the complexity of political regimes can be understood, in the short term, as analogue to decentralizing them. As discussed in previous chapters, the network approaches of new public management point to decentralization as a reality. Thereby, it appears that there is already a tendency in regimes to complexify.

One way to accelerate the process would be by thinking which dynamics in political regimes could facilitate sociopolitical self-organization, for instance, the life-like properties[73] intrinsic to human social systems, and extrapolate them to the structures, rules and dynamics of political regimes. Political regimes could also incorporate into their formal mechanics the self-assemble and self-disassemble of other structures and rules coming as positive feedback loops from human social systems. These two properties could make that the interactions between the elements of political regimes and human social system would shape the structure, rules and, ultimately, the dynamics of regimes without resorting to any formal procedure or without limiting the elements that interact in them. Basically, it is important to think in non-fixed topologies and in the absence of supreme laws of general application, such as constitutions, for reducing the chances of emergence of violent vias for opposing formal top-down institutions.

---

[71] Compare Kauffman. *At Home in the Universe: The Search for the Laws of Self-Organization and Complexity.*

[72] The definitions were taken from Rescher, Nicholas. *Complexity: A Philosophical Overview.* 1989.

[73] Self-*properties are certain characteristics, behaviors or features that living systems and life-like systems present without the need of any external help. The descriptions were taken from Lordache, Octavian. *Self-Evolvable Systems.* 2012. And Mitchell, Melanie. "Complex Systems: Network Thinking". 2006. Electronic document.

**Table 2. Modes of Complexity.**

| Modes of Complexity | | |
|---|---|---|
| Compositional Complexity | **Constitutional Complexity** | Number of constituent elements or components |
| | **Taxonomical Complexity** | Variety of constituent elements |
| Structural Complexity | **Organizational Complexity** | Variety of different possible ways of arranging components in different modes of interrelationship |
| | **Hierachical Complexity** | Organizational disaggregation into subsystems |
| Formulaic Complexity | **Descriptive Complexity** | lenght of the account that must be given to provide an adequate description of the lenght of the system. |
| | **Generative Complexity** | Lenght of the set of instructions that must be given to provide a recipe for producing the system at issue |
| | **Computational Complexity** | Amount of time and effort involved in resolving a problem |
| Functional Complexity | **Operational Complexity** | Variety of modes of operation or types of functioning |
| | **Nomic Complexity** | Elaborateness and intricacy of the laws of governing the phenomena at issue |

Source: Author´s own elaboration.

Additionally, these structures, dynamics and rules could self-configure and self-reconfigure, adjusting the parameters or geometry of the regime and modifying its behavior when required or desired. In few words, it would be important to think in global patterns that emerge out of local interactions without any global coordination -not even leadership. The fact that each node and subnetwork of political regimes could decide which links and subnetworks form or break, immediately, opens the door for who can be part of decision-making processes of political systems, i.e., it could bring down the barrier that currently exists between the civil society and the political society, pushing them closer to becoming one single networked entity. However, the latter does not necessarily mean the absence of basic principles for self-organizing human interactions.

Self-organization in living organisms uses constrains in the form of positive and negative feedback loops. Translated into political regimes this could be expressed in local arrangements that rearrange and eliminate failing nodal interactions and develop protection and immune mechanisms from the bottom-up, avoiding reappearances of equal or similar nature of the failures, and preventing that particular connections or clusters generate failure cascades -only with the use of local information. It would be very interesting for the combination of the civil and the political society, if increasing the importance of direct citizen´s participation mechanisms, would generate an emergent pattern capable of self-diagnosing, self-repairing and self-healing thanks to its decentralized dynamics that provide local information. Having simple basic protocols for interacting socio-politically would make the structures, rules and dynamics resulting from the interplay of human social systems and political regimes more robust because they would function over adaptive principles. Thus, robustness, with its foundations on flexibility, could make coherent patterns to emerge. Coherent patterns would imply that disruption would not be the result of the dynamism of the environment. Nevertheless, for this scenario to be possible, regimes need to open to new types of nodal connections, structures, rules and dynamics. Augmenting their modes of complexity points exactly in that direction.

The idea is that, after breaking the barrier, political regimes could consider sociopolitical self-organization as valid as their traditional mechanisms of organization. And despite that regimes and human social systems would not behave as a sole collectiveness of nodes or as a single network, it would be interesting to generate interdependent dynamics that could increase their network consciousness (being aware that local interactions can have chaotic effects).

It is highly plausible that with this interplay between semi-formal institutions and self-organized dynamics, in the long term, the top-down control mechanisms of organization that political regimes use and their tree topologies will be replaced by structures, rules and dynamics that instead of being previously fixed in an organizational chart or constitution, would be dynamic, evolvable and, specially, synthetized from the bottom-up.

The self-*properties of living and life-like systems translated to the structures, rules and dynamics of political regimes would not counteract the natural tendency towards self-organization of human social systems. Instead, it would bring better possibilities for finding optimal solutions to the problems that political systems face when they compute the inputs of human social systems. Self-organization as a reality in complex systems validates the absence of ruling authorizes in the life-like systems that human social systems are.

### 3.2. SELF-ORGANIZING HUMAN SOCIOPOLITICAL INTERACTIONS

This section discusses two books that have studied self-organization in living systems, *At Home in the Universe*[74], by Stuart Kauffman and *Self-Organization in Biological Systems*[75], by Scott Camanzine et al., and extrapolates their concepts and theories about self-organization to the study of how human social systems should self-organize their sociopolitical interactions and how sociopolitical self-organization is possible, plausible and preferable.

Stuart Kauffman is a theoretical biologist that has dedicated the last decades to the study of the laws of complexity and self-organization. He proposed the idea of *order for free,* which stresses that under certain parameters in -non-linear- systems composed by many elements, self-organized patterns naturally emerge[76]. Patterns refer to "an organized arrangement of objects in space and time".[77] They include global structures, rules or dynamics resulting from interactions that (self-) organize without any external intervention. But patterns can also arise from defined, predictable sources. For instance, in contemporary world the interplay of rigid political regimes imposing order upon human social systems makes nation-states and the concept of citizenship to emerge as global patterns.

Coherent and organized patterns in complex systems emerge non-teleologically from non-fixed interactions. The nation state and being a citizen are previously fixed patterns, since individuals cannot decide whether to be included or not in their flows and dynamics. Nonetheless, following the definition of patterns in complex systems, nation-states and being an obligated citizen are not precisely organized emergencies. The reason is that the complexity of systems is attributed to the nonlinear emergencies they produce. So, when global patterns are expected in complex systems, the interactions that produce them become a closed end in itself that block the natural flow of self-organization that could have make something truly organized to arise. In complex systems, pattern formation should be achieved without the intervention of any outsider, as Camanzine et al claim. Yet, political regimes (the political society, external entities) look to organize human social systems (the civil society) to form nation-states by means of top-down control.

[74] Compare Kauffman, Stuart. *At Home in the Universe: The Search for the Laws of Self-Organization and Complexity.* 1995.
[75] Compare Camanzine et.al. *Self-Organization in Biological Systems.*
[76] Compare Kauffman. *At Home in the Universe: The Search for the Laws of Self-Organization and Complexity.*
[77] See Camanzine et.al. *Self-Organization in Biological Systems.* p.8.

Self-organization is possible in complex systems with large number of elements that share common properties, like being part of a species or community -despite individual diversity and heterogeneity. Camanzine et al. use fish schools for explaining why self-organized patterns are natural emergences in these type of systems, but the idea can be extrapolated to human social systems:

> […] in schools or [communities] containing thousands [or millions] of fish [or humans], it is inconceivable either that one supervisory individual [president, king, emperor, etc.] could monitor everybody´s position and broadcast the moment-by-moment instructions needed to maintain the school´s [nation-state´s] spatial structure, or that individual fish within the school [citizens] could monitor the movements of the leader and follow accordingly. Coherence is achieved, instead, by each fish gathering information only about its nearest neighbors and responding accordingly.[78]

The question that rises is why, if humans present more complexity than fish, political regimes insist on organizing them with the use of leaders and top-down control. Undoubtedly, the best via would lie on information gathered in local interactions of individuals or groups, i.e., self-organization. Indeed, Camanzine et al. claim that "one of the mayor problems associated with having a complex system run by a central authority is that it requires both an effective communication network among individuals and sophisticate cognitive abilities by the central planner"[79].

Self-organized –obviously, decentralized- coordination is, indeed, a better alternative in complex systems, given that no individual in a complex system, not even leaders, can have complete knowledge about everything that is occurring in them at every moment. Camanzine et al. claim that decision-making processes in social animals involve so many interactions that are non-intuitive. Clearly, an argument against organizing human social systems in a top-down fashion by means of tree topologies with leaders that selectively transform inputs into decisions.

---

[78] See Camanzine et.al. *Self-Organization in Biological Systems.* p.22.
[79] See Camanzine et.al. *Self-Organization in Biological Systems.* p.64.

Stuart Kauffman favors democracy[80] as "the best way to solve complex problems"[81] (in human social systems). However, without noticing it, he provides a very good argument against democracy when he refers to how space states in complex systems are too huge to be calculated. The space state conveys all the possible set of behaviors that a system can adopt. For the case of human social systems, their space state is so immense – or hyperastronimical, as Kauffman names them- that trying to define one and only one by means of a central authority is absurd. Especially because systems with hyperastronimical possible space states are extremely unpredictable.

As a mental experiment to illustrate the problem addressed in this monograph, lets imagine a nation-state conformed by N=3 communities, P, Q and R. Each community is formed by N´=50 individuals. Each individual interacts with others inside their own communities, but, at the same time, they interact with individuals of other communities. They can form groups at meso-levels and interact as groups, even outside of their community. Communities can interact between them and they can also form groups and interact with other groups of communities too, but they can also interact with groups and individuals as well. Figure 13 illustrates this example.

[80] It is assumed that he refers to welfare-states and representative democracy, since it is the predominant nowadays.

[81] See Kauffman. *At Home in the Universe: The Search for the Laws of Self-Organization and Complexity.* p.28.

**Figure 13. Interactions at micro, meso and macro-levels.**

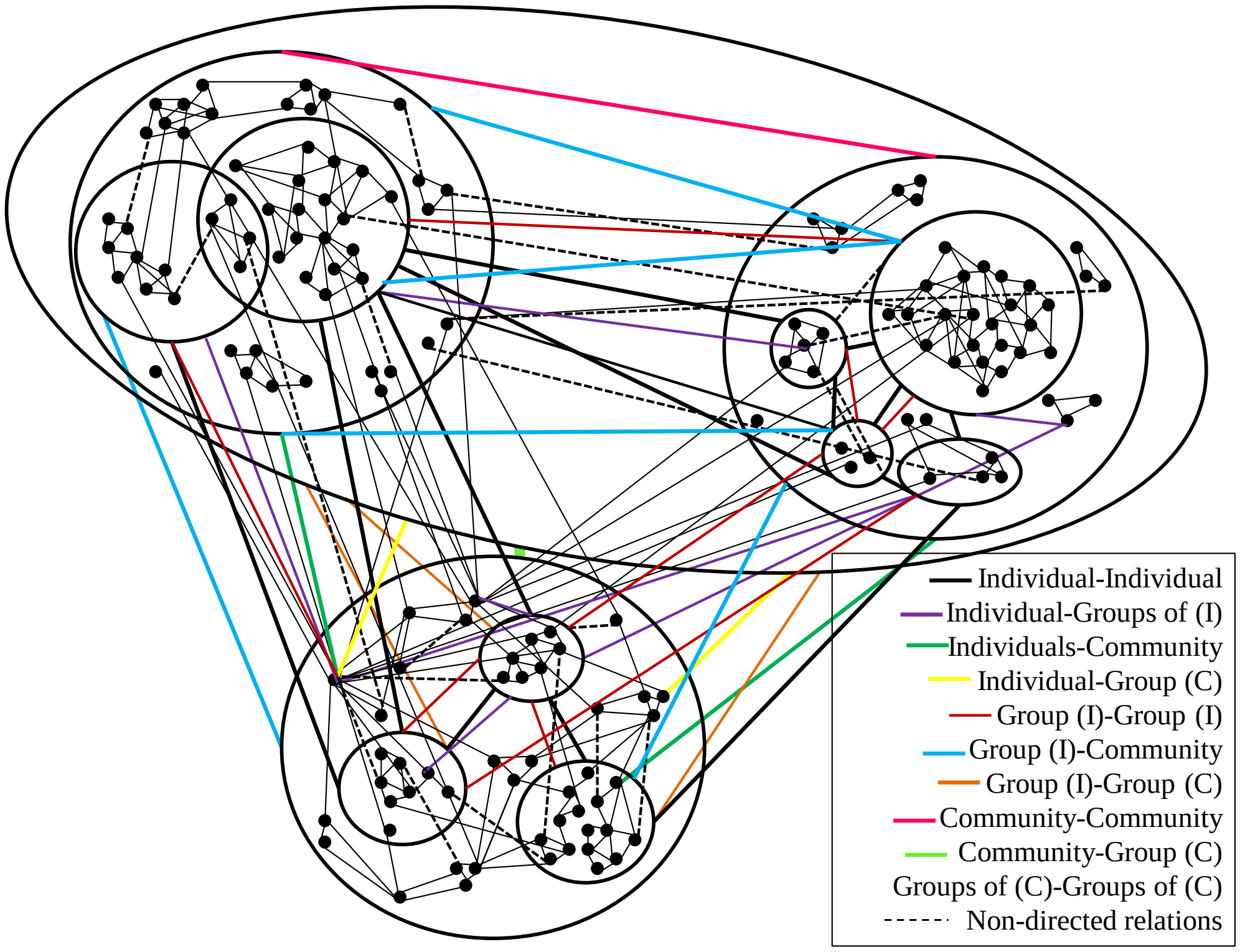

Source: Author´s own elaboration.

Now, let us dramatically reduce the complexity of the interactions of our theoretical nation-state and imagine that each individual only interacts with individuals in their communities in the form of the basic structure of human interactions (figure 10, directed triangles) and that the latter are isolated, meaning that the behavior of each individual is influenced by inputs coming only from 2 other individuals, and that this is the *unique* way in which each individual interact. If we take Kauffman´s space states and calculate the value for each theoretical community, the result would be defined by $2^{50}$. That is, 1.125.899.906.842.624 possible forms of relations, only considering this hyper-simplistic example of real life. Hence, it is inevitable to question: which optimization procedure based on deliberation processes between humans selected by means of influence and popularity within a organizational network structure shaped as a tree topology for favoring the formal computational dynamics of modern democracies can handle such number of space states?. In other words, which leader or representatives can be aware of such hyperastronimical number of possibilities in order to choose the best?

Besides leaders, Camanzine et al. speak of alternatives sources of order to self-organization, such as blueprints, recipes and templates. According to their descriptions of the latter, templates can be assimilated as a constitution because they guide the definition of the pattern. Blueprints resemble laws because they specify what should be built -what type of society we want. And recipes, the "sequential instructions that precisely specify the spatial and temporal action of the individual´s contribution to the whole pattern"[82], assertively define what codes or statutes are. There are also, of course, external forces that shape the top-down organization of human social systems such as international treaties and agreements. The point of this analogy relies on stressing how the organization of human social systems is tried to be achieved step-by-step by, basically, every possible alternative to self-organization: leaders, blueprints, templates and recipes. At the same time. The main reason for preferring self-organization to the latter alternatives is how improbable it is for leaders, recipes, blueprints and templates to coordinate so many interactions and elements in human social systems.

Another negative aspect of the alternatives to self-organization is how they regulate interactions ex-post using an evolutionary speed that is not synchronized with the velocity of changes in human social systems. As an example, contemporary political regimes regulate homosexual relations millennia after they started. This is not very coherent. If sociopolitical interactions were self-organized, it would be possible to implement the previously mentioned basic protocols of interactions open to changes in social systems at the moment when they occur. Even better, this protocols do not need even to formally exist.

For instance, in a global system composed by communities P, Q and R, each interaction or subnetwork within them could have its own values, principles and basic protocols that provide positive feedback loops while working as attractors in the interactions between communities. Positive feedback loops, as Kauffman states, are the main responsible for making systems to change, adapt and evolve, while negative feedbacks try to keep the systems in their current states or pull them back to their original one by avoiding or reducing the effects of fluctuations. Kauffman stresses that, in complex environments, systems must be stable enough to support random fluctuations, but not too static that any minimal fluctuation would make the system to collapse. This is the kind of environment where sociopolitical self-organization can prosper. Basic protocols in self-organized interactions could produce coherent, yet flexible, arrangements in an interplay between positive and negative feedbacks that may lead to the emergence of organized global patterns in human social systems. It would be preferable, of course, if these were bioethically-grounded and had only a minimum level of complexity. They could combine and give rise to other protocols, open to new diversities and that also co-evolve as responding differently to the tuning their parameters.

[82] See Camanzine et.al. *Self-Organization in Biological Systems*. p.49.

Camanzine et al. mention that complex systems must have tunable parameters, from where their flexibility emerges. It is valid to think which could be the ones of human social systems that could allow them to evolve their own dynamics of (self-) organization. Even, maybe, institutions, but without institutionalism becoming an obstacle for the emergence of organized patterns. Without doubt, this corresponds better with the complex nature of human social systems than leaders, blueprints, recipes and templates.

Self-organized sociopolitical interactions would be a mutation in cultural evolution that may simplify the rules of interactions in human social systems. In that way, new diversity in the latter would not need to enter political systems as inputs in order to be legitimated. A simple data base of basic protocols, some of them working as attractors, yet open to transformations and, mostly, to new emergencies, would anticipate the incorporation of more diversity and complexity to the social systems. Just like natural selection has made with genomic information in living organisms[83]. Because although the behavior of complex systems can be coupled by attractors that can produce order in very large systems, this works, mostly, if they are small, thus, negative feedbacks in the simple rules of interactions (such as the respect for life and life´s dignity) can sustain the stability needed to self-organize complex human social systems in even more complex and dynamic environments. Indeed, self-organization in dynamical environments requires both, positive and negative feedback loops, to maintain the internal coherence of the systems and to allow perturbations, changes and harmonic fluctuations to take place.

There are two types of perturbations that can affect systems (in this case, political regimes): internal and external. Political regimes have such intrinsic duality when responding to internal and external perturbations, that their structures and rules are designed for strengthening the barrier between the political and the civil society. For instance, if an external perturbation changes the global pattern of political regimes, they behave as chaotic systems producing dysfunctional behaviors –especially if the top-node is attacked –or killed. On the contrary, when an internal perturbation changes a pattern in the regime, it behaves as a completely organized system because, in many cases, the system remains the same. In this case, negative feedback outweighs positive feedbacks because regimes respond as completely ordered systems. The latter case can be understood better thinking about internal cascades of corruptions, which are usually ignored for a longue period outside of political regimes. This duality in responses makes the organization of political regimes unreliable because it divides more the civil and the political society, inhibiting the possibility of human social systems to self-organize. The reason is that positive feedback loops coming from individuals outside the regimes are barely integrated into the global formation of the pattern, unless they formally enter political systems. This is why political regimes need to increase their structural, compositional, functional and operational complexity for individuals to participate directly in local decision-making processes.

[83] Compare Camanzine et.al. *Self-Organization in Biological Systems.*

Kauffman points out that one feature that controls if attractors make systems to behave as organized systems, chaotic or at the edge of chaos is having defined basic protocols for the functioning of the elements interacting. Fundamentally, rules defined by *and* ($\wedge$) and *or* ($\vee$) relations produce order. The rest may lead to chaotic dynamics. This is an argument against the current systems of laws in political regimes, which get too complicated by adding new laws and codes and statutes, some of them derogating others, replacing them, invalidating them, etc. Maybe, systems of rules defined by 1s and 0s would facilitate the interactions of human social systems at every scale. They could easily be coded in the data bases of the protocols underlining the dynamics of the sociopolitical and computational processes. This may function similarly to Internet, which has basic classical topologies as its core and, yet, it is the most complex artificial system invented to date. This can lead towards better communication dynamics that result in more coherent global patterns.

An interesting way to solve problems related to the overlapping of rules that organize human social systems could be eliminating the artificial boundaries of States, triggering cascades of self-organized diversity through networks of migration, technology, trade, culture, etc. This would increase the complexity of human interactions until they reach a point where hierarchies -and their capacity to organize social systems in a top-down fashion- would be reduced to a non-elegant feature of the past.

Kauffman attributes the emergence of order in systems that co-exist within complex environments to collective dynamics of networks that arise spontaneously. At the same time, he attributes self-organization to evolution by natural selection –as well as Camanzine et al. Communities of humans should use this knowledge to turn towards human social systems that manage to generate organized patterns from the self-organization of their sociopolitical interactions, as a sophisticated evolution in cultural selection that could counteract the many disadvantages that regimes with tree topologies imply. It is very probable that the self-organization of sociopolitical interactions makes clusters of individuals (communities) and clusters of communities to emerge. As interactions would not be previously fixed, they may be formed by preferential attachment. This is why complex network structures may be the emergent pattern of sociopolitical self-organization.

Complex networks have a rich diversity of subnetworks that corresponds with the community diversity of human social systems. Biological organisms provide the best inspiration for organizing human social systems because the latter are complex systems and present life-like properties. But, most importantly, because living organisms have been solving optimization problems for millions of years[84], many of them related to their structural and dynamical optimization. –the main concerns of decision-making processes in political systems.

In sum, the self-organization of human sociopolitical interactions should be allowed so that coherent organized global patterns emerge. A good way to encourage sociopolitical self-organization would be ending with the barrier between the political and the civil society. Political regimes do not need to impose upon or top-down control human social systems to organize them, because they already have their own internal dynamics of

---

[84]See Casti, John. "Biologizing Control Theory: How to make a Control System Come Alive". *Complexity.* Vol.7, No.4 (2002). p.10.

self-*control, due to self-organization. Decentralized control is, by far, the best way to organize complex systems, but, even there, control should come from internal interactions, instead of imposed by any external entity. This is why augmenting the number, scopes and importance of direct citizen participation mechanisms would be a suitable alternative for triggering the decentralization of political regimes and, ultimately, sociopolitical self-organization. Although transforming political regimes from the inside by having the diversity of human social systems as an excuse would also help to augment their complexity. Whatever is used to do so, any self-organized via would be much better than top-down traditional forms of organizing social systems. Including the veiled mechanisms of coercion of representative democracies such as elections and voting dynamics.

# 4. HOW CLASSICAL POLITICAL REGIMES INHIBIT THE SELF-ORGANIZATION OF HUMAN SOCIAL SYSTEMS. TOP-DOWN CONTROL THROUGHOUT HISTORY

This chapter illustrates how classical political regimes are far from properly organizing human social systems or allowing the latter to self-organize. The basic power structures of classical political regimes that shaped western´s political history are illustrated and some of the disadvantages of organizing human social systems by means of top-down control are explored. Additionally, it is speculated which could be the emergent global pattern of sociopolitical self-organization.

Throughout history, political regimes have had structures, rules and dynamics whose properties block and inhibit the emergence of sociopolitical self-organization –at least without being equivalent to rebellion, insurrection or subversion or provided with the need of entering the computation of political systems as inputs. One reason for this is that regimes have never been structured outside of the guidelines of tree topologies and their top-down mechanisms of control, despite the apparent structural and dynamical changes that have occurred along the centuries in terms of moving among various political systems. In reality, as it will be exemplified, the latter does not imply any real transformation regarding the way in which political regimes try to organize human social systems: by imposing control upon them.

For instance, western classical political history started with the Greeks and the polis around the fifth century BCA. Two of the most important Greek polis were Athens and Sparta. Both organized human social systems with top-down control -feasible to be described by tree topologies. Sparta´s architecture (figure 14b), an aristocratic-monarchy, had Spartans at the top node of the topology directing the transmission of information along the core channel. Some groups under the yoke of Spartans could not intervene at all in political life. Athens (figure 14a) was a democracy. It had an assembly of citizens (the *ecclesia*) with a daily-changed president and, in an inferior level, there were some magistrates. Citizens of Athens gathered around to discuss and decide about Athens's organization in the public arena (the *agora*). However, women, men under 18, slaves, foreigners and men without military training were not citizens, thus, they could not participate in decision-making processes. They were simply top-down organized. The Roman civilization is another milestone of political history. Rome underwent many different faces. It was a monarchy, a republic and an empire[85]. The three periods obey to top-down control modes of organizing human social systems, as figures 14c, 14d, 14e illustrate. The fall of the Roman Empire left profound voids in power-holding, economy direction and cultural crises that lead territories to experience a privatization of political power in the hands of feudal lords (counts, dukes and princes), who were, at the same time, subordinated to the authority of the king, who had very strong political relations with the church. Non-secular political power was typical of the middle-ages, in as much that it sustained the separated social classes of Feudalism, commonly represented with pyramids similar to figure 14f.

Basically, in the middle-ages there were polyarchic governments (monarchs) and religious authorities with administrative and jurisdictional aptitudes[86]. At the beginning of modern political thought, monarchies unified the feudalist world of medieval politics in the figure of national states. States became the new higher authority, in replace of god[87], occupying the main direction of the organization of human social systems under their control.

Thomas Hobbes in his book *Leviathan*[88] supports top-down mechanisms for organizing human social systems. Hobbes proposed the idea of states as big entities with the right of imposing themselves upon human social systems by means of the monopoly of violence. For him, the absence of coercive regimes would give rise to what he named *the state of nature*[89], a disorganized stage of human social systems previous to the ruling of states –clearly, Hobbes was not aware of the virtues of self-organization.

In the eighteen century, the United States of America promulgated the first constitution where the supreme authority was not a monarch but an elected president, instituting the modern base for contemporary representative democracies[90]. John Stuart

---

[85] Compare Frost Abbott, Frank. *A History and Description of Roman Political Institutions.* 2001.
[86] Compare Hernández-Becerra, Augusto. *Las Ideas Políticas en la Historia.* 2008.
[87] Compare too Bakunin, Mijail. *God and the State.* 2008.
[88] Compare too Hobbes, Thomas. *Leviathan.* 1994.
[89] A stadium before, tacitly, accepting being ruled by states.
[90] Compare Collier, Christopher. *Decision in Philadelphia: The Constitutional Convention of 1787* . 1986.

Mill[91] lived during that time and despite that for him the ideal form of government was representative democracy (figure 14g), he saw that regimes could easily turn into a majority tyranny or in a system that reproduces exploitation patterns towards the mass by the minority that holds power and properties –and he was right. It could be said that representative democracies do not organize human social systems with top-down control because governors are elected bottom-up. Moreover, that in such regimes decisions are not implemented in a top-down fashion since they are backed by citizens and policy formation is an integral process. However the civil society has very restricted vias for acting as positive feedback loops to the authoritarian decisions that result from decision-making processes, in comparison to the ones of the political society.

Dictatorships are taken to practice trough political regimes with only one individual occupying the node on top of the structure (figure 14h). Dictatorships have a rigid core channel and strict control mechanisms. Usually, only one party or group holds political power and orders have to be top-down complied, whether individuals belong or not to the topology or only to the social systems under its control.

Welfare States started to develop in the form of representative democracies after the Cold-War, many of them with prime ministers, parliaments, councils or cameras (figure 14i). Presumably, these States focus on guaranteeing social rights to individuals through the creation of specific institutions for doing so and by increasing the influence of syndicates, citizens´ participation mechanisms, and guarantied public services, health, education, among others. Also, in principle, they allow positive feedback loops coming from the civil society. In the twenty-first century many supranational institutions and cross-national organizations were created for guarantying such rights. Nonetheless, there is still a longue path to cover before this can be completely accomplished. Sociopolitical self-organization can reduce the time.

Today, the world is still state-centric, but the tendency is to lowering the importance of states as the local revitalizes at global scales[92] giving rise to more networked structures that emerge from local interactions. The scopes of sociopolitical self-organization will not be complete until institutionalism cease to exist as the only way in which sociopolitical interactions are validated. Because historically, political thought has emphasized on the need of governments and formal institutions, lowering the importance of self-organized sociopolitical interactions taking place in human social systems or outside political regimes, augmenting the chances of emergence of actors that with the wrong mechanisms try to be included in the computational processes of political systems. In few words, political regimes are the main cause of disorder and political violence around the world because governments ignore that the institutionalization of politics (*politiké*) is just one aspect of political dimensions.

---

[91] Compare Mill, John Stuart. *Considerations on Represetative Democracy.* 1991.

[92] Compare United Citied and Local Governments (UCLG), *Descentralization and Local Democracy in the World: First Global Report 2008.* 2008.

Resorting to the beginning of western classical political thought, the Greeks were aware that politics extends beyond institutions to a more social framework, where it is the common world a society builds together, independent to one form a form of government or another –or to the absence of one. This is politics as *politéia*, which refers to cooperation and consensus and includes all ethical, administrative, social, philosophical, educational, scientific, aesthetical and religious aspects of societies, as *Carlos Eduardo Maldonado*[93] states. Therefore, the term sociopolitical encompasses, basically, every aspect of human life that can be subject of been seen politically (in the sense of politéia). As a matter of fact, sociopolitical self-organization could be the best expression of a *politéia.*

[93] Compare Maldonado, Carlos Eduardo. "Política y Sistemas No Lineales: la Biopolítica". In *Dilemas de la Política.* 2007.

**Figure 14. Top-down control throughout history.**

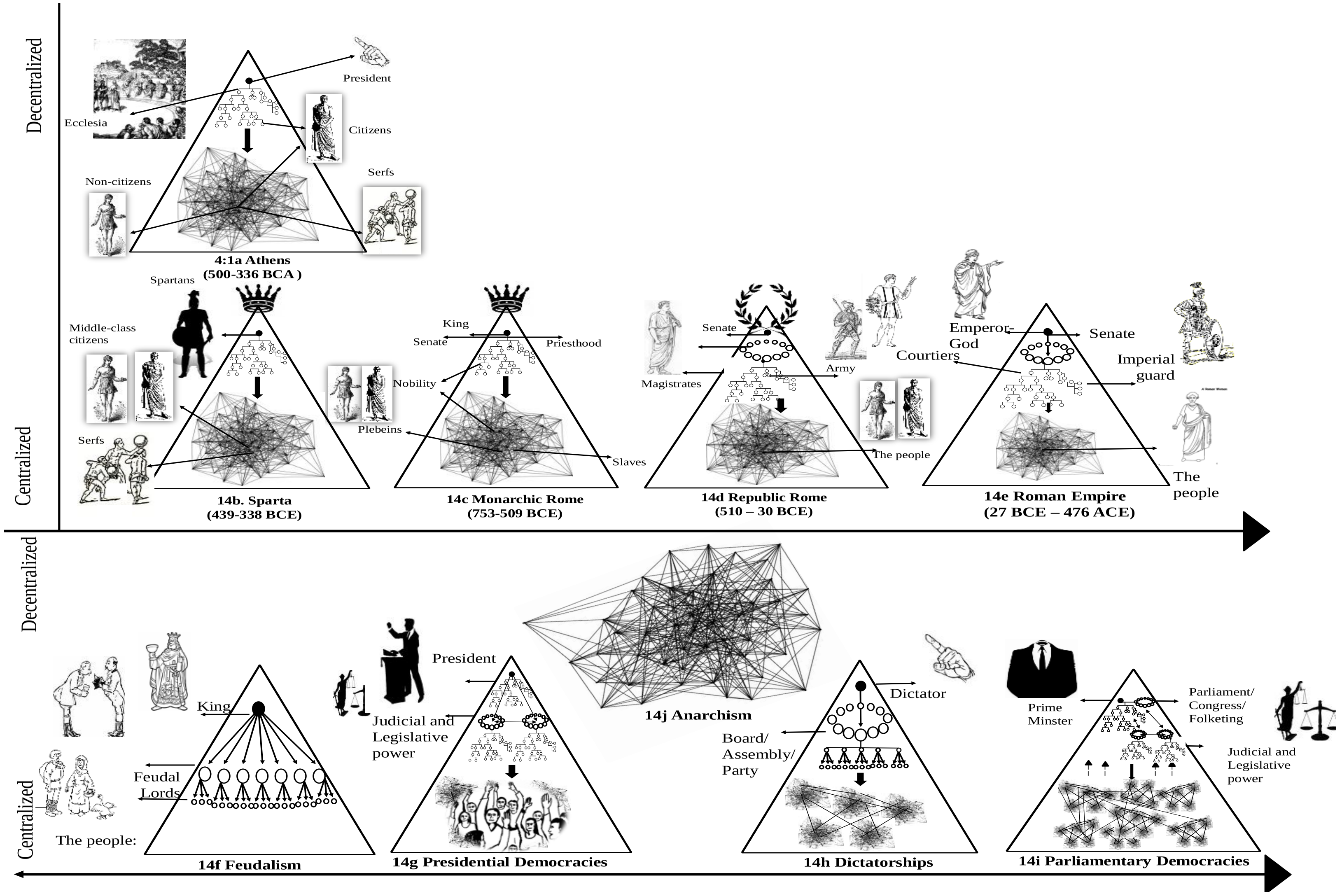

Source: Author´s own elaboration.

Besides libertarian socialists there also exist authoritarian socialists. What distances them from anarchists is that the latter do not sympathize with top-down control and neither with the idea of governors. Karl Marx[94] and Friedrich Engels[95] were authoritarian socialists and two of the fathers of communism. They made a very similar critique to capitalist modes of production, as the one that John Stuart Mill made to representative democracy. Authoritarian socialists did not agree with economic power being held in only few hands and wanted people to revolt against the rich minority, in order to replace them. In respect to political regimes, they claimed that a communist economic system would conduct states to disappear because there would not exist anymore a class to oppress, since the majority of individuals, the proletariat, would hold some type of power. Mijail Bakunin, another of the greatest exposures of Anarchism, agreed with their *critique*. For Bakunin, states create and guaranty the permanent existence of a governmental aristocracy opposed to the people. However, Bakunin claimed that the dictatorship of the proletariat would organize human social systems, anyway, by means of a top-down power structure -with the slightly difference that the nature of the individuals on top would change[96]. Bakunin´s critique became a reality when the soviets led the Bolsheviks to power and they became a closed and unique party ruling the Soviet Union for some decades[97] and when the communist Cuban revolution led towards a one-party democracy. Serge Galam explains that this happened because communist organizations are grounded over democratic centralism, "which is nothing else than a tree-like hierarchy".[98]

In short, political regimes have varied along the centuries but no profound transformation in the way in which human social systems are organized has taken place. The only changes that have occurred relate to subtleties like how individuals get or are placed at the top nodes of the topology –but, mostly, who they are. This means that history has always relapsed on structures, rules and dynamics that do not reflect the complexity of human social systems and that constraint sociopolitical self-organization.

---

[94] Compare Marx, Karl. *Capital.* 1995.
[95] Compare Engels, Frederick. *The Origin of the Family, Private Property and the State.* 1972.
[96] Compare Shatz, Marshal. *Bakunin, Statism and Anarchy.* 2002.
[97] Compare Guerin, Daniel. *L´Anarchisme, De la Doctrine a L´Action.* 1965.
[98] See Galam, Serge. "Democratic Voting in Hierarchical Structures or How to Build a Dictatorship" *Advances in Complex Systems.* Vol.3, No.1-4 (2000).p.76.

The anarchist proposal is the closest that human social systems can get to the self-organization of their sociopolitical interactions, given that their structural and dynamical properties are very close to each other[99] (figure 15), yet, anarchism has always been pejoratively seen. Nevertheless, it is hoped that complexologists realize how the sciences of complexity, the study of complex systems and the theory of self-organization inevitably lead towards anarchic forms of sociopolitical organization and interactions.

**Figure 15. Anarchy and social networks.**

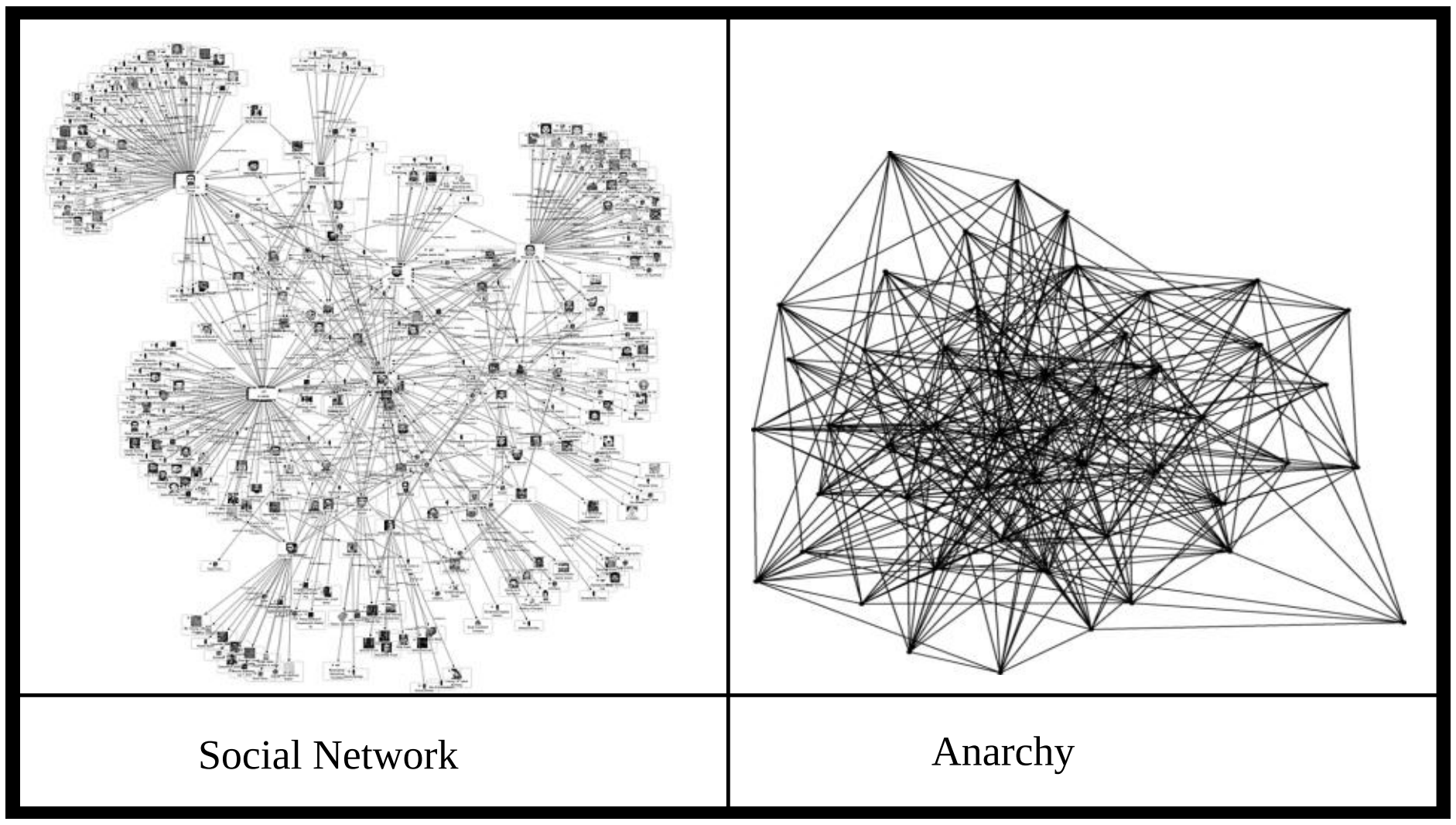

Source: Author´s own elaboration.

Many disadvantages can be named in relation to organizing human social systems in a top-down fashion. For instance, being part of the political society becomes an end in itself because the individuals and groups occupying the nodes on top of the topology or those directly linked with the core channel have comparative benefits in relation the rest nodes of the topology but, especially, in comparison to those individuals and groups who are not part of the topology at all and are ruled by it (the civil society). This increases the barrier that there is between the political society and the latter, closing even more political regimes and inhibiting self-organized sociopolitical interactions to prosper. In fact, when sociopolitical self-organization develops outside of the closed boundaries of political regimes, it does not have as much importance as the sociopolitical interactions formally computed by political regimes. In other words, "political authorities tend to generate increasing return processes"[100] because the power of political regimes self-reinforces, augmenting the capabilities in life of the nodes on top of the structure at the expense of the human social systems that they organize.[101]. In sum, sociopolitical self-organization would imply better capabilities for human social systems than the ones allowed by political regimes, their top-down control and their tree topologies, since inputs would not be processed selectively.

[99] The topology of the social network was reproduced from Advanced Systems Group. Social Network Analysis. Sentinel Visualizer. Electronic document.

Another disadvantage of organizing human social systems in a top-down fashion is the vulnerability of the structure used to do so, i.e., tree topologies. Human social systems have complex, diverse, noisy and nonlinear sociopolitical dynamics before entering political systems. And when political regimes try to organize them by implementing the outputs of political systems through political regimes with tree topologies, their closed mechanisms remain short. Political regimes need structures capable of not being overwhelmed by the complexity of human social systems because their rigidness makes them greatly vulnerable. Classical political regimes lack the adaptive behaviors that make a system resilient. For instance, a regime structured as a tree topology can collapse if the core channel or top node is affected or taken down -because all the strength of the architecture depends on its power core.

Additionally, tree topologies are not the most suitable structure for information processing in a complex system. It is impossible for a political regime structured as a tree to be aware of the huge amount of information flowing in human social systems. Self-organized sociopolitical interactions would not present this disadvantage, since they only require the use of information gathered locally for making organized patterns to emerge.

Because of their top-down organization and their control hierarchies, it is not very probable that innovation, the engine of cultural evolution, emerges, easily, in classical political regimes. Control hierarchies imposes barriers to innovation[102] because they base upon principles of causality and the assumption of linear interactions, which leaves no room for emergence –and closes the door for self-organization.

---

[100] See Ma, Shun-Yun. "Political Science at the Edge of Chaos? The Paradigmatic Implications of Historical Institutionalism" . (2007). p.61.
[101] Capability refers to the possibilities and opportunities of individuals for mobilizing resources and transform their environments and contextual conditions into ones that allow them to happily and harmonically exist within them: Compare Sen, Amartya. *Development as Freedom.* 1999.
[102] Compare Abbott, Russ. "Putting Complex Systems to Work". *Complexity.* Vol.13, No.2 (2007).

Classical political regimes try to set the dynamics of each node of the regime and of human social systems. And although regimes admit that human social systems change, they use *ex-post* formal mechanisms -such as laws and normativity- that incorporate changes too slowly, in comparison with the velocity of evolution of human social systems. Co-evolutionary networks of cooperation, labor specialization and trade are a natural emergence of network exchanges in human social systems[103] when they are not being constrained by an external entity. Thereby, it would be better not to impose barriers to their self-organization by means of regimes with tree topologies that block them more than what they allow them to evolve.

In addition to the latter, "topology has a strong influence on coalition emergence".[104] And because the tree topologies operate in a top-down fashion, the chances of presenting failure cascades (figure 16), for instance, in the form of public corruption increases. Although it is probable that some form of corruption can also emerge in self-organized sociopolitical interactions, the impact would be much lower, since self-organization operates with local exchanges of information. Thus, despite possible chaotic effects, they would not affect every dimension of human social system.

Complex networks –also called scale free networks- are highly robust and resilient. Statistically, they can resist random attacks because there is no such thing as a core channel or a top node with comparative greater importance than the others. This means that failures in one node, connection or group of them do not necessary affect negatively the performance of the complete structure.

---

[103] Compare Hayek, Friedrich. *Law, Legislation, and Liberty. Vol. 2.* 1973; Compare Ridley, Matt. *The Rational Optimist: How Prosperoty Evolves.* 2010.

[104] Compare Norman-Salazar et.al. "Emerging Cooperation on Complex Networks" *Proceedings of the 10th International Conference on Autonomous Systems.* (2011)

**Figure 16. Corruption cascades in tree topologies.**

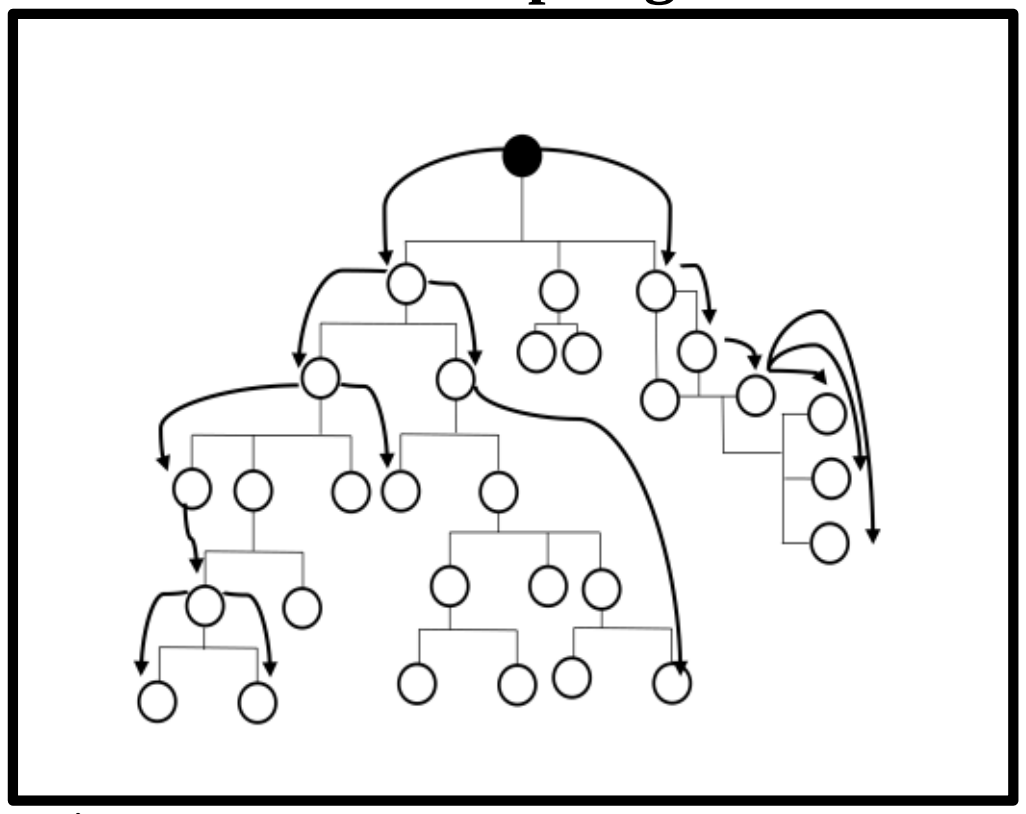

Source: Author´s own elaboration.

Top-down power structures increase the inequality among individuals and social systems instead of looking to equally beneficing them[105]. However, humans ruled by tree topologies are not the only ones who are negatively affected by top-down organization. Given that human social systems are organized starting from global ideas about them, their environments can also be profoundly harmed because local information about their ecosystems and its bio-diversity is usually ignored. Organizing social systems in a top-down fashion goes against protecting the dynamic equilibrium of life on Planet Earth.

Summarily, regimes with tree topologies are very restricted, oppose to the complexity of human social systems[106], to the self-organization of their sociopolitical interactions and their control hierarchies barely promote positive feedback loops coming from lower-levels of the structure. Trees may be suitably for local telecommunication networks and LAN design, but they are not an appropriate topology when it comes to the organization of human social systems for the many disadvantages that they imply for human social systems.

The disadvantages mentioned above show that self-organized sociopolitical interactions have always needed mechanisms inside political regimes to be legitimated, thus, sociopolitical self-organization has never been a reality. Nevertheless, it is important to consider that *now* is the best moment to start, consciously, increasing the complexity of political regimes because contemporary world is moving away from the time where tree topologies could, until certain extent, frame human sociopolitical interactions without possible chaotic effects. It is also important to consider that human social systems are being complexified at an accelerating range as microelectronics-based technology increases its role in sociopolitical development[107] -like Internet and mobile devices. If a change is not addressed now, then the effects of the strengthened interdependence among human social systems may overturn against their own evolution. The idea of superior -and geographically-based- political regimes through which legitimate sociopolitical dynamics goes against the complexity of contemporary world.

[105] Compare Hsu, Sara. "The Effect of Political Regimes on Inequality", 2008. Electronic source.
[106] Compare Banathy, Bela H. *Designing Social Systems in a Changing World.* 1996.

In addition to this, there are non-microelectronics-based technological advances (and computational systems) that will influence future decision-making dynamics in human social systems with scopes that have never been considered before this monograph, such as computation using biological materials, computation inspired on the functioning of living systems, living technology, protocells, bio-inspired artificial intelligence, organic computing and biological hypercomputation[108]. Hence, top-down organization should be replaced for structures that flow in harmony with the cultural and technological transformations that occur in human social systems.

For these reasons, political regimes should avoid structures, rules and dynamics that prevent them from reflecting the complexity of the social systems they try to organize. Top-down control expressed in the form of coercive, hierarchical, pyramidal and linear arrangements by no means recalls the complex nature or the tendency towards self-organization that human social systems present.

---

[107] Compare Castells, Manuel, y Gustavo Cardoso. "The Network Society: From Knowledge to Policy" In *The Network Society: From Knowledge to Policy, 2*005.

[108] This idea developed in conversations with Nelson Alfonso Gómez Cruz and is subject of a work in progress. The complete list of the biologically-based fields that will influence the future of political systems computational processes is developed in Gómez-Cruz, Nelson Alfonso. *Computación y Vida.* Bogotá: Desde Abajo, 2013.

# 5. SELF-ORGANIZED HUMAN SOCIAL SYSTEMS AND POLITICAL REGIMES WITH COMPLEX NETWORK STRUCTURES: THE EMERGENCE OF ANARCHY

Up until now it has not been directly stated that the background idea of this monograph is, indeed, an anarchist proposal based on the properties and behaviors of complex systems. As stated in the last chapters, anarchic sociopolitical interactions have the only topology and self-organized dynamics that can reflect the complex nature of the structures and dynamics of human social systems. Therefore, self-organized sociopolitical interactions would be so decentralized that, basically, they can only result in govern-less emergent patterns.

Complex networks are the kind of topology provided with the dynamics, diversity, complexity and organization that might emerge from decentralizing political regimes, after merging them with human social systems and subsequently to increasing the scopes of sociopolitical self-organization and complexifying political regimes. Nevertheless, reaching a point where a political regime co-exists with self-organized sociopolitical interactions equals to making regimes vanish because no institution can handle such complexity. Basically, sociopolitical self-organization in a world where politics is understood as a *politéia* would imply as much elements and interactions as the ones in human social systems. This can only be equivalent to an anarchic sociopolitical global pattern.

As shown in figure 10, the basic structure of self-organized social relationships between humans is a non-directed triangle. Within a sociopolitical context, this triangles joined together would form similar structures to figure 17. Additionally, links in sociopolitical self-organization between individuals would be very dynamic because nodes could create or break connections without restrictions. Therefore, in self-organized sociopolitical interactions each node would choose where to attach following individual preferences, optimization and randomness –which is how complex networks are formed[109]. In this process, trees may suffer transitions similar to the hybrid networks in next figure, as they would lose their hierarchical nature and become more heterarchical in mezzo-levels, before making anarchical global patterns to emerge.

[109] Compare Barabási, Albert-László. "Network Science: Luck or Reason". Nature 489 (2012): 507-508.

**Figure 17. Hybrid networks resulting from augmenting the complexity of political regimes.**

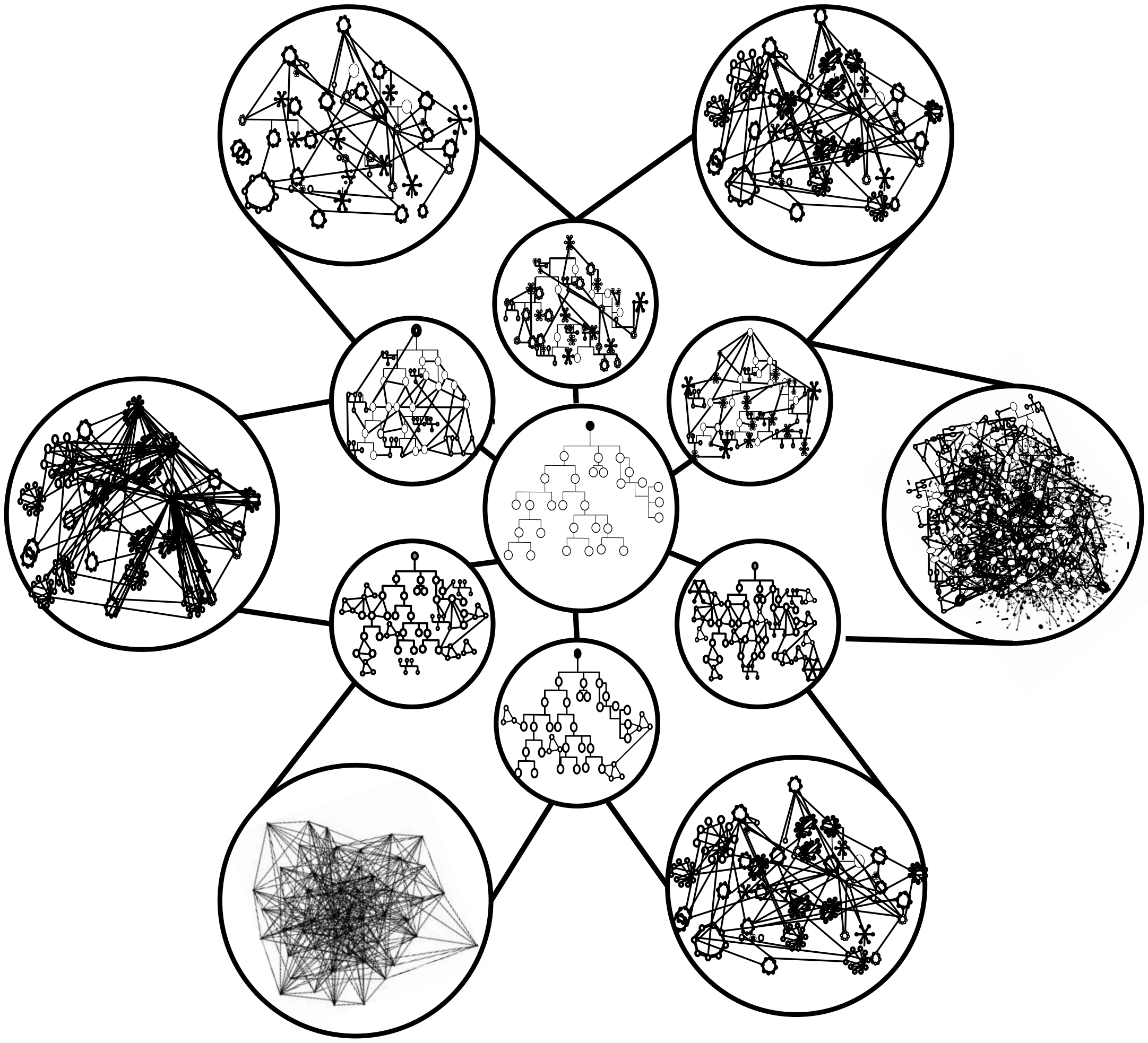

Source: Author´s own elaboration.

In the middle of trees and complex networks there would be many other types of networks. Since each one, at least in principle, would be self-organized, this situation can lead towards local optimal computational performances in decision-making processes. And despite that preferential attachment implies that "nodes with higher degree (of connections) receive more new links than nodes with lower degrees"[110], this does not means that the nodes that used to be at the top of the tree would be the ones with more connections[111] - since the connectivity of the emergent topology would obey a dynamical topology that would not be built over top-down dynamics. The possibility, however, cannot be discarded.

[110] Compare Mitchell, Melanie. "Complex Systems: Network Thinking".

[111] For instance, webpages like Google tend to generate more connections in comparison to less connected webpages, but there is no subordination relation between Google and the webpages attached to it.

Connections grown by preferential attachment are feasible to be described by a power-law and structured as a complex network. It is highly probable that this global structure presents self-similar and fractal patterns because self-similarity is a characteristic of networks grown by preferential attachment.

Self-similarity is a property of fractals[112] that refers to how a *part* of a system resembles the global structure, but in a smaller scale. Self-similarity can be statistical (table 3)[113], as in natural fractals, or exact, as in mathematical fractals (table 4)[114]. The type of self-similarity that may arise with the self-organization of human sociopolitical interactions will be statistical (table 5b). However, it does not only refer to the emergent structure but also to the rules and dynamics: among different scales of the complex network, functions, subnetworks, rules and dynamics could, but not necessarily, be repeated iteratively.

**Table 3. Natural fractals: statistical. Self-similarity.**

| | | | |
|---|---|---|---|
| 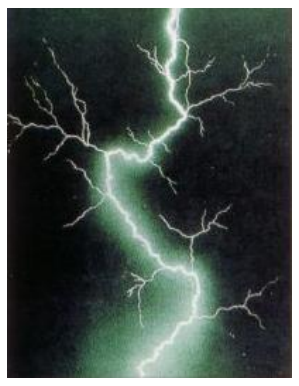 | 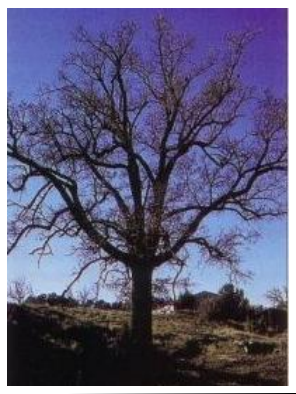 | 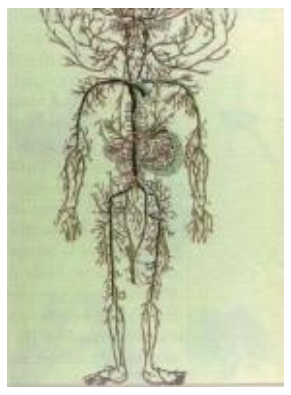 | 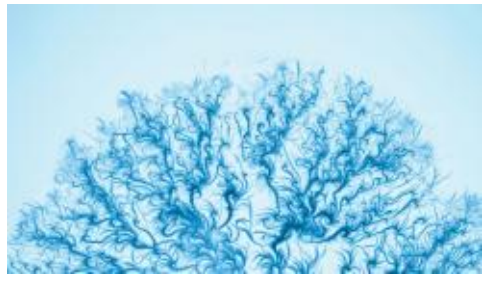 |
| 3a Lightning | 3b Top of a tree | 3c Circulatory sysrtem | 3d Bacteria colony |
| 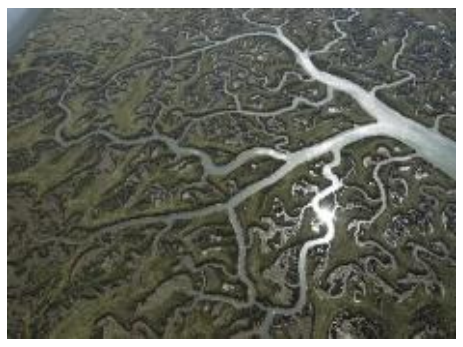 | 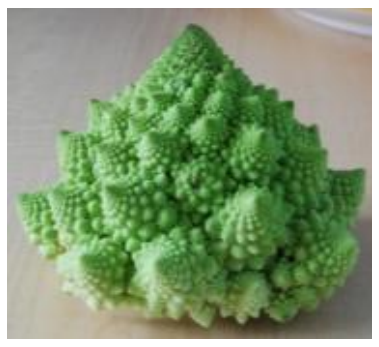 | 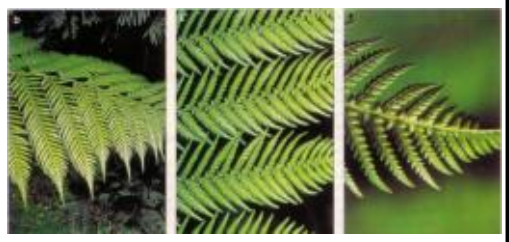 | 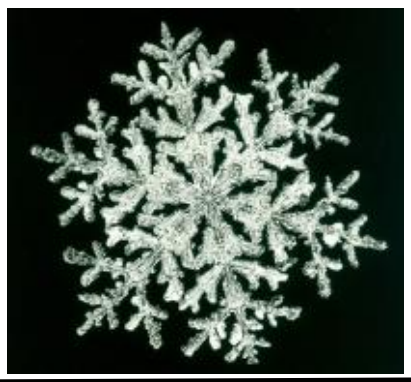 |
| 3e River with tributaries | 3f Romanesco coliflower | 3g Fern | 3h Snowflake |

Source: Author´s own elaboration.

[112] Fractal geometry is the geometry of natural and living structures, discovered by the mathematician Benoît Mandelbrot (Mandelbrot, Benoît. *La Geometría Fractal de la Naturaleza.* 1977) who unified his findings about general patterns in natural and living structures, giving birth to the geometric study of nature. Natural and living forms, such as clouds, mountains, seacoasts and lightings cannot be properly be described in terms of Euclidean Geometry, i.e., the geometry of circles, triangles, squares, cubes and pyramids. The irregular structures of fractals adds difficulty to the task of locating them into one of the three spatial dimensions of Euclidian Geometry. Indeed, Mandelbrot coined the name fractal after the latin word *fractus*, which translates irregular. Understanding fractality is important for the development of this idea because complex networks present self-similar patterns, and self-similarity is a distinctive mark of fractal structures.

[113] "Lightning". Electronic document; "Tree". Electronic document; "Bacteria colony" Electronic document; "River and tributaries"; "Romanesco coliflower". Electronic document; "Fern". Electronic document; "Snowflake". Electronic document.

[114] "Sierpinski triangle". Electronic document; Koch snowflake". Electronic document; Mandelbrot set". Electronic document; Fractal fern; Fibonacci fractal". Electronic document; Julia set". Electronic document.

**Table 4. Computer-generated fractals: exact self-similarity.**

| | | |
|---|---|---|
| 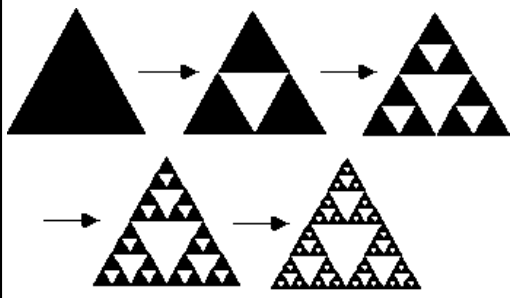 | 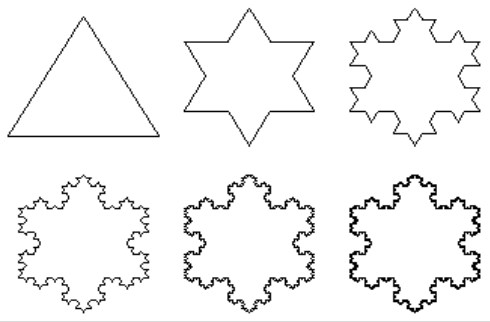 | 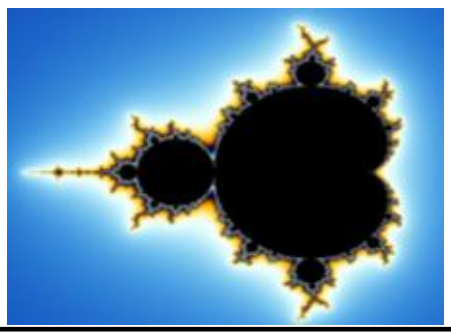 |
| 4a step by step iterated Sierpinski Triangle | 4b Koch Snowflake | 4c Mandelbrot set |
| 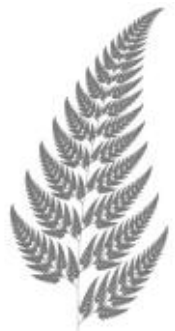 | 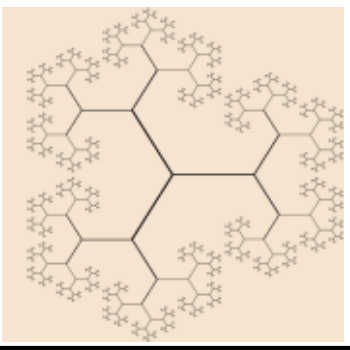 | 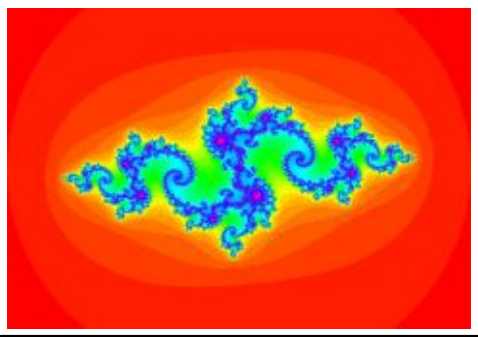 |
| 4d Fractal Fern | 4e Fibonacci Fractal | 4f Julia Set |

Source: Author´s own elaboration.

Summarily, with time, hybrid networks that present fractal patterns (table-5b) and grown by preferential attachment will continue to non-linearize their self-organized dynamics in such way that they will give rise to complex network structures[115] (table 5c). Complex networks resulting from sociopolitical self-organization, in principle, would allow every individual of human social systems to participate directly in decision-making processes in various networks at the time, eliminating the necessity of delegating participation in the name of representatives that, in most cases, are not very representative.

Complex networks also imply better computational performances in decision-making processes because the information used in the interactions that generate them are local, which reduces noise, failures and short views in decision-making processes, a very common situation that takes place when global views are assumed. This could lead towards better ways of organizing human social systems than when regimes use top-down control, leaders, templates and recipes, pretending to have a global view about them.

[115] Compare Barabási, "Network Science: Luck or Reason". (2012).

**Table 5: Preferential attachment and statistical self-similarity preceding complex networks.**

| | | |
|---|---|---|
| 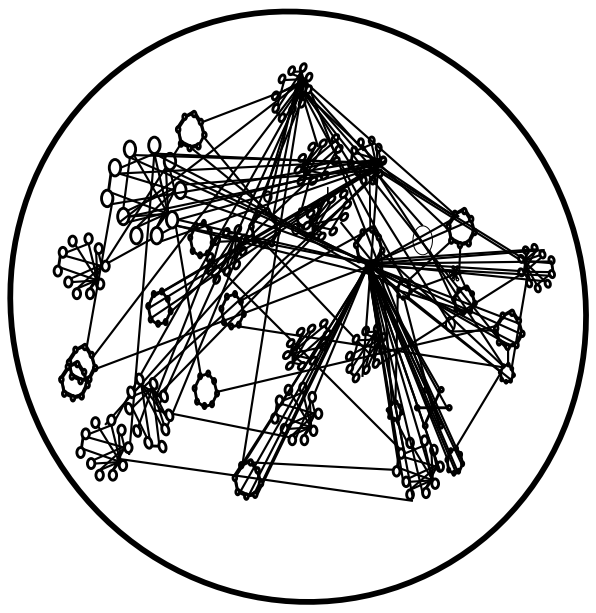 | 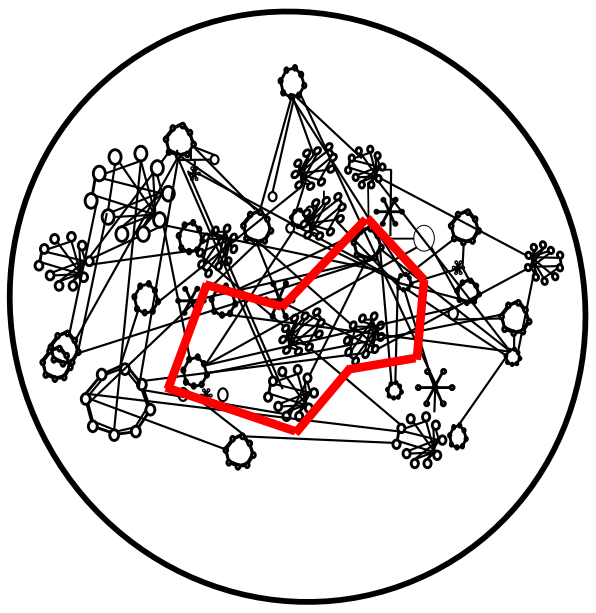 | 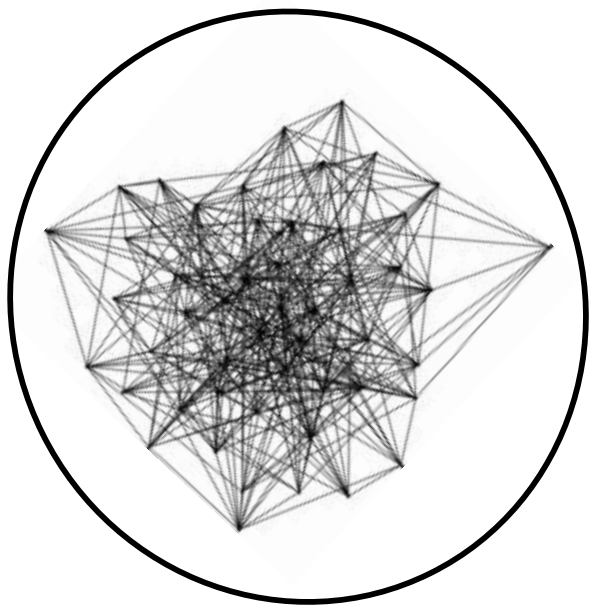 |
| 5a Preferential attachment | 5b. Statistical self-similarity | 5c. Complex network |

Source: Author´s own elaboration.

The self-organization that would make anarchy and complex networks to emerge (anarchic complex networks) can lead towards the disappearing of political regimes. In part, the transition could take place undercovered and from their inside. This may not be as fast as wished because it implies to have the current states of the world as the initial conditions[116]. However, as Kauffman[117] states, random fluctuations can be amplified by positive feedback-loops, independently of the latter, meaning that starting from trees to form complex networks by means of increasing the complexity of the dynamics of the first (and combining them with sociopolitical self-organization) would be no hindrance. Anyway, the emergent pattern in human social systems would be more adaptive and robust than actual ones because it could continue working under different parameter ranges given that complex networks present life-like properties. Thus, they are very adaptive. Nevertheless, individuals at the top nodes of political regimes and the political society, in general, do not want regimes to augment their complexity –and neither too many decentralized dynamics; thus, non-radical transformations may be the most peaceful via towards letting human social systems self-organize, for preventing that the later use their capabilities in behalf of not losing the advantages of being positioned in the tree topologies.

[116] Nonetheless, this monograph bases on a *longue durée* approach: Braudel, Fernand. “Histoire et Sciences Sociales: La longue durée ”. *Annales H.S.C.* Vol.13, No.4. 1958.
study, i.e., specific events are mostly discarded.

[117] Compare Kauffman, Stuart. *At Home in the Universe: The Search for the Laws of Self-Organization and Complexity*.

Complex network structures as the emergent pattern of sociopolitical self-organization would be so vast, diverse, heterogeneous and boundary-less that they would be human social systems self-controlling, self-evolving, self-configuring, self-reconfiguring and self-disarranging thanks to the networked interdependence of their self-organized interactions. It is possible, of course, that hierarchies emerge in some subnetworks, but the most probable is that they may respond to scalar hierarchical principles, and not precisely to control hierarchies. And if they do, the positive feedback of the basic protocols and the dynamism of sociopolitical self-organization might lead those subnetworks to break, rearrange and reconfigure into more organic arrangements.

Self-organized human social systems would not be contained under any rigid structure because heterarchical networks of networks of individuals, groups, communities, etc. will prevail. It is claimed here that this would bring more peaceful interactions between individuals, groups, communities and social systems in general, since there would not exist any superior establishment to enroll, pervade, cooptate, permeate, dwindle or knock out in order to participate in decision-making processes. The new interactions of what used to be classical political regimes (figure 5) would mutate to interactions that correspond with figure 18.

**Figure 18. Interactions in complexified political regimes.**

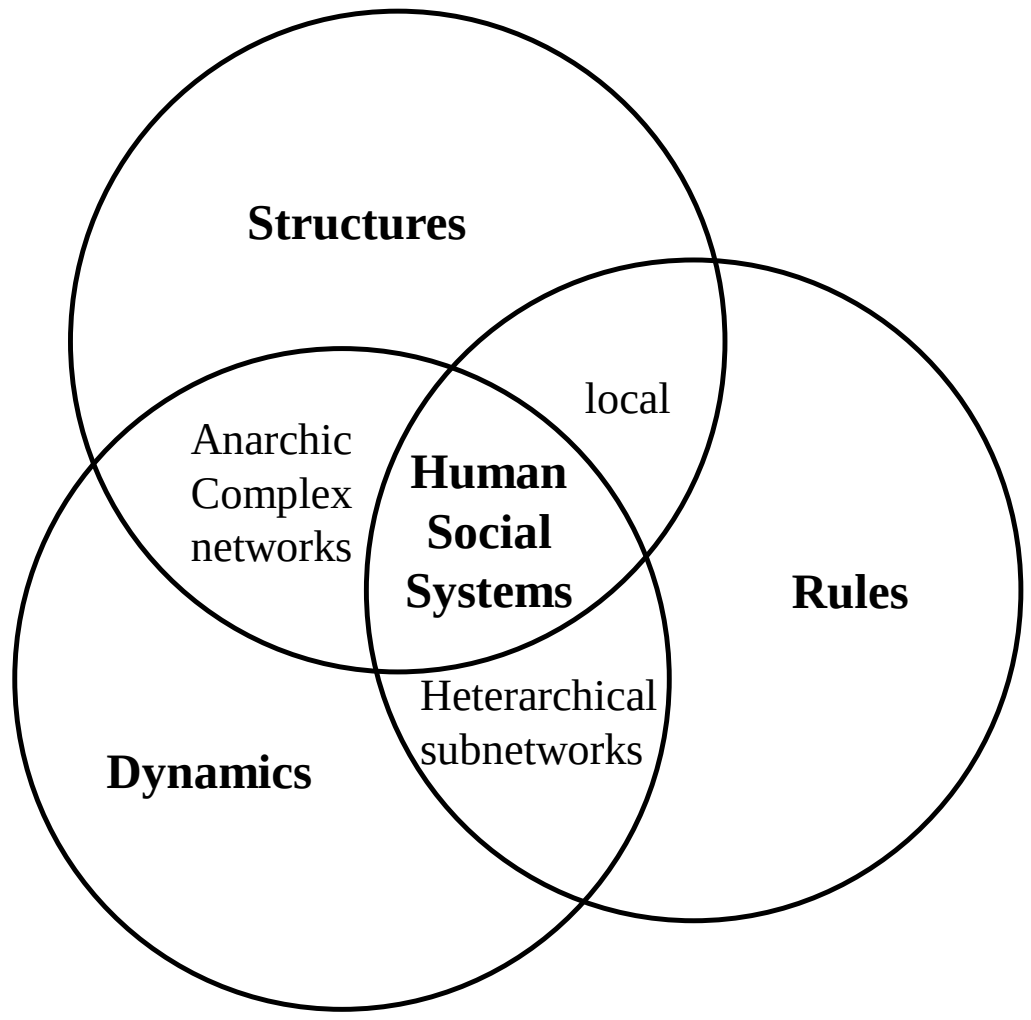

Source: Author´s own elaboration.

Political regimes are considering more and more dynamics of information flow (diversity, inclusion and migration) as inputs of political systems. Maybe, without even being aware of its implications in the long term -regarding an augment in the complexity of regimes and in the possibility of human social systems to self-organize. Additionally, if the impact of Internet over human sociopolitical interactions is also considered, it might be that a silent dwindling of top-down political power structures is already happening. Anarchy may already be emerging. Reaching a point where the complexity of political regimes has increased so much that they loss their capacity to frame the sociopolitical interactions of human social systems appears to be the tendency. This may be the last century of political regimes as we know them. The idea of anarchic complex networks comes from this loss in the capacities of central controllers to organize human social systems.

Ultimately, anarchic emergent patterns generated from sociopolitical self-organization and structured as complex networks would imply that the outrages historically committed towards groups of individuals in the name of preserving an institutions may not be possible anymore. Despite that political science fears anarchy, when seem from a complexity sciences perspective, anarchy (particularly, anarcho-communism) entails swarm-like self-organized sociopolitical interactions that elegantly co-evolve between evolution and entropy in spaces of flowing harmony.

# 6. CONCLUDING REMARKS AND FUTURE WORK

In this thesis I summarized the structural and dynamical history of classical political regimes and I came to the conclusion that there has not been any profound structural or dynamic transformation in the way in which political regimes organize human social systems -despite historical changes in political systems. This means that we have been relapsing for the past two and a half thousand years into the same topology, but with different names. I presented some of the disadvantages of having political regimes that organize human social systems in a top-down fashion and stated that the best way in which human social systems could be organized was by means of their own self-organization. I proposed that one via to get closer to the sociopolitical self-organization of human social systems was breaking the barrier between the political and the civil society. However, for this transition to take place harmonically, I suggested that sociopolitical self-organization could be pursued undercover by means of augmenting the complexity of political regimes, or decentralizing them, and opening the spectrum of who could be part of decision-making processes, for instance, increasing the influence and diversity of citizen´s direct participation mechansms. I claimed that, in doing so, political regimes and human social systems could become a sole entity in the long term. Complexified political regimes (with hybrid structures, basic protocols and complex dynamics) may formed by connections that, after a while, would follow by preferential attachment and would start to present self-similarity, getting closer to resembling the structure of human social systems. I stressed that complex networks would be the emergent global pattern of sociopolitical self-organization. And because preferential attachment emerges from optimization and randomness, I pointed out how self-organized sociopolitical interactions would allow optimal computational performances in decision-making processes. As information gathered would be local and would not follow any centralized mechanism of coercion, it is highly probable that the resulting global pattern would be anarchic and would entail more peaceful or, at least, harmonic sociopolitical flows. This was the antiestablishment undercover statement of the monograph, because if regimes complexify and their influence decay in behalf of the self-organization of human social systems, political regimes as we know them would finally disappear, since there would be no difference between the synthetized dynamics of organization in human social systems and the latter.

In this monograph I proposed an idea. My next step is to test it. With the use of evolutionary algorithms and agent-based modeling, in my next level of education I intend to prove whether the self-organization of human sociopolitical dynamics is optimal for decision-making processes in human social systems and if sociopolitical self-organization would result in global complex network patterns. Whatever it is that I find, I am deeply convinced that the top-down control we have been using for organizing sociopolitical interactions is not an adequate way for addressing the complex problems of our social systems. Desperately, our world needs more organic structures.

# BIBLIOGRAPHY

Bakunin, Mijail. *God and the State*. New York: Cosimo, 2008.

Bakunin, Mikhail. *Selected Writings From Mikhail Bakunin: Essays on Anarchism*. St. Petesburg, Florida: Red and Black Publishers, 2012.

Banathy, Bela H. *Designing Social Systems in a Changing World*. New York: Plenum Press, 1996.

Barabási, Albert-László. *Linked: How Everything Is Connected to Everything Else and What It Means Means Business, Science and Everyday Life*. New York: Plume, 2003.

Barclay, Harold. *People without Government: An Anthropology of Anarchy*. London: Kahn & Averill Publishers, 1996.

Bertalanffy, Ludwig Von. *General System Theory: Foundations, Development, Applications*. New York: George Braziller Inc, 1969.

Camazine (et al). *Self-Organization in Biological Systems*. Princeton, New Jersey: Princeton University Press, 2001.

Collier, Christopher. *Decision in Philadelphia: The Constitutional Convention of 1787*. New York: Ramdom House, 1986.

Easton, David, *Esquema para el Análisis Político*. Buenos Aires: Amorrortu, 2006.

Engels, Frederick. *The Origin of the Family, Private Property and the State.* New York: Pathfinder Press, 1972.

Estrada, Ernesto. *The Structure of Complex Networks*. Oxford: Oxford University Press, 2012.

Fairtlough, Gerard. *The Three Ways of Getting Things Done*. Axminster: Triarchy Press, 2006.

Forgacs, David. *The Antonio Gramsci Reader. Selected Writings 1916 - 1935*. New York: New York University Press, 2000.

Frost Abbott, Frank. *A History and Description of Roman Political Institutions.* Boston: Adamant Media Corporation, 2001.

Goldsmith, Stephen, y William, D. Eggers. *Governing by Network*. Washington, D.C: The Brookings Institution, 2004.

Gómez Cruz, Nelson Alfonso. *Computación y Vid*a. Bogotá: Desde Abajo, 2013.

Guerin, Daniel. L´Anarchisme- *De la Doctrine a L´Action*. Paris: Gallimard, 1965.

Haken, Hermann. *Information and Self-Organization: A Macroscopic Approach to Complex Systems*. Berlin: Springer, 2006.

Hayek, Friedrich. *Law, Legislation, and Liberty. Vol. 2.* Chicago: University of Chicago Press, 1973.

Hernández-Becerra, Augusto. *Las Ideas Políticas en la Historia*. Bogotá: Universidad Externado de Colombia, 2008.

Hobbes, Thomas. *Leviathan.* Indianapolis: Hackett Publishing Company, 1994.

Kauffman, Stuart. *At Home in the Universe: The Search for the Laws of Self-Organization and Complexity*. New York: Oxford University Press, 1995.

Knoke, David. Political Networks: *The Structural Perspective*. New York: Cambridge University Press, 1990.

Kropotkin, Piotr. *Mutual Aid: A Factor of Evolution. Montreal*: Black Rose Books, 1996.

Lapierre, Jean-William. *El Análisis de los Sistemas Políticos*. Barcelona: Península, 1976.

Lordache, Octavian. *Self-Evolvable Systems*. Berlin: Springer, 2012.

Maldonado, Carlos Eduardo. *Complejidad: Revolución Científica y Teoría*. Bogotá: Universidad del Rosario, 2009.

Maldonado, Carlos Eduardo, y Nelson Alfonso Gómez Cruz. *El Mundo de las Ciencias de la Complejidad*. Bogotá: Universidad del Rosario, 2011.

Mandelbrot, Benoît. *La Geometría Fractal de la Naturaleza*. Barcelona: Tusquets, 1977.

Marx, Karl. *Capital*. Oxford: Oxford University Press, 1995.

Mill, John Stuart. *Considerations on Representative Democracy*. New York: Prometheus Books, 1991.

Mitchel, Melanie. *Complexity, A Guided Tour*. Oxford: Oxford University Press, 2009.

Mitleton-Kelly, Eve, (ed). *Complex Systems and Evolutionary Perspectives of Organizations: The Application of Complexity Theory to Organizations*. Oxford: Elsiever Science, 2003.

Morin, Edgar. *El Método I, La Naturaleza de la Naturaleza*. Madrid: Cátedra, 2006.

Morin, Edgar. *El Método II, La Vida de la Vida*. Madrid: Cátedra, 2006.

Morin, Edgar. *El Método III, El Conocimiento del Conocimiento*. Madrid: Cátedra, 2006.

Morin, Edgar. *El Método IV, Las Ideas.* Madrid: Cátedra, 2006.

Morin, Edgar. *El Método V, La Humanidad de la Humanidad, La Identidad Humana.* Madrid: Cátedra, 2006.

Morin, Edgar. *El Método VI, Ética*. Madrid: Cátedra, 2006.

North, Michael, y Charles M. Macal. *Managing Business Complexity*. London: Oxford University Press, 2007.

Oppenheimer, Priscila. *Top-Down Network Design*. Indiaapolis: Cisco Press, 2004.

Prigogine, Ilya. *Thermodynamic of Irreversible Processes*. New York: John Wiley & Sons, 1955.

Rescher, Nicholas. *Complexity: A Philosophical Overview*. New Jersey: Transaction Publishers, 1989.

Ridley, Matt. *The Rational Optimist: How Prosperity Evolves*. New York: Harper Collins, 2010.

Sen, Amartya. *Development as Freedom*. New York: Random House, 1999.

Shatz, Marshal. Bakunin, Statism and Anarchy. Cambridge: Cambridge University Press, 2002.

Watts, Duncan. *Six Degrees. The Science of a Connected Age*. New York: Norton & Company, 2003.

Woodcock, George. *Anarchism. A History of Libertarian Ideas and Movements*. New York: The World Publishing Company, 1962.

Book Chapters

Castells, Manuel, y Gustavo Cardoso. "The Network Society: From Knowledge to Policy". In The Network Society: From Knowledge to Policy, 3-21. Washington D.C.: John Hopkins University, 2005.

Maldonado, Carlos Eduardo. "Política y Sistemas No Lineales: la Biopolítica". In B. Vela Obregón. Dilemas de la Política, 91-142. Bogotá: Universidad Externado de Colombia, 2007.

Journal Papers

Abbott, Russ. "Putting Complex Systems to Work". Complexity Vol. 13, No. 2 (2007): 30-49.

Barabási, Albert-László. "Network Science: Luck or Reason". Nature 489 (2012): 507-508.

Bodin, örjan, y Jon Norberg. "Information Network Topologies for Enhanced Local Adaptive Management". Environmental Management Vol. 35, No. 2 (2005): 175-193.

Bogason, Peter, y Julieth A. Musso. "The Democratic Prospects of Network Governance". The American Review of Public Administration Vol. 36, No. 3 (2006): 3-18.

Braudel, Fernand. "Histoire et Sciences Sociales: La longue durée". Annales. Historie, Sciences Sociales. Vol. 13. 4 (Oct-Dec 1958): 725-753.

Cano, Pedro (et al). "Topology of Music Recomendation Networks". Chaos Vol. 16 (2006): 013107 1 - 013107 3.

Casti, John L. "Biologizing Control Theory: How to make a Control System Come Alive". Complexity 7, No. 4 (2002): 10-12.

Colomer, Joseph M. "Desequilibrium Institutions and Pluralist Democracy". Journal of Theoretical Politics, Vol. 13, No. 3. 2001: 235-247.

Crumley, Carole L. "Heterarchy and the Analysis of Complex Societies" Archeological Papers of the American Anthropological Association. Special Issue: Heterarchy and the Analysis of Complex Societies Vol. 6, No. 1 (1995): 1-5.

Cudworth, Erika, y Stephen Hobden. "Anarchy and Anarchism: Towards a Theory of Complex International Systems". Milennium-Journal of International Studies Vol. 39 (2010): 399-416.

Easton, David. "An Approach to the Analysis of Political Systems" World Politics Vol. 9, No. 3 (April 1957): 383.

Elgie, Robert. "The Classification of Democratic Regime Types: Conceptual Ambiguity and Contestable Assumptions". European Journal of Political Research, 1998: 219-238.

Fraley, Dennis J. "Ubiquity Symposium "What is Computation?": Computation is Process". Ubiquity, November (2010): 1-6.

Fuchs, Christian. The Political System as a Self-Organizing Information System. In Trappl, Robert. Cybernetics and Systems, Vol.1. Vienna: Austrian Society for Cybernetic Studies, 2004. 353-358.

Galam, Serge. "Democratic Voting in Hierarchical Structures or How to Build a Dictatorship". Advances in Complex Systems Vol. 3, No. 1-4 (2000): 171-180.

Gemmill, Gary, y Charles Smith. "A Dissipative Structure Model of Organization Transformation" Human Relations Vol. 38, No. 8 (1985): 751-766.

Gordon, Deborah. "Dynamics of Task Switching in Harvester Ants". Animal Behavior. Vol.38, No.2 (1989): 194-204.

Kapucu, Naim. "Interagency Communication Networks During Emergencies" American Review of Public Administration. Vol.36, No.2 (2006): 207-225.

Klijn, Erik-Has. "Analyzing and Managing Policy Processes in Complex Networks: A Theoretical Examination of the Concept Policy Network and its Problems". Administration & Society, 1996: 90-119.

Kuchaiev, Oleksii (et al). "Topological Network Alingment Uncovers Biological Function and Phylogeny". Interface, 2010: 1-14.

Lewin, Robert, Teresa Parker, y Birute Regine. "Complexity Theory and the Organization: Beyond the Metaphor". Complexity Vol. 3, No.4 (1998): 36-40.

Ma, Shun-Yun. "Political Science at the Edge of Chaos? The Paradigmatic Implications of Historical Institutionalism". International Poliical Science Review. Vol.28, No.1 (2007): 57-78.

Maldonado, Carlos Eduardo. "La Complejidad es un Problema, no una Cosmovisión". UCM Revista de Investigación Vol. 13 (2009): 42-54.

McCulloh, Warren, S. "A Heterarchy of Values Determined by the Topology of Nervous Nets". Bull. Math. Biophysics Vol. 7 (1945): 89-93.

Meek, Jack W, Joe De Ladurantey, y William H. Newell. "Complex Systems, Governance and Policy". Emergence: Complexity and Organizations Vol. 1, No.2 (2007): 24-36.

Morçöl, Göktuğ, y Aroon Wachhaus. "Network and Complexity Theories: A Comparison and Prospects for a Synthesis". Administrative Theory & Practice Vol. 31, No.1 (2009): 44-58.

Norman-Salazar, Juan A. (et al). "Emerging Cooperation on Complex Networks". Proceedings of the 10th International Conference on Autonomous Systems. Taipei: Tumer, Yolum, Sonenberg and Stone (Eds), 2011. 669-676.

O´ Toole, Laurence. "Shaping Formal Networks Through the Regulatory Processes". Administration and Society Vol. 36, No. 2 (2004): 186-207.

O´Toole, Laurence J., y Kenneth J. Meier. "Desperately Seeking Selznick: Cooptation and the Dark Side of Public Management in Networks". Public Administration Review Vol. 64, No.6 (2004): 681-693.

Pappi, Franz Urban, y Christian H.C.A. Henning. "Policy Networks: More than a Metaphor?". Journal of Theoretical Politics Vol. 10 (1998): 553-575.

Prettejohn, Brenton; Methew J Berryman, and Mark D. McDonnell. "Methods for Generating Complex Networks with Selected Structural Properties for Simulations: a review and tutorial for neuroscientists". Frontiers in Computational Neuroscience Vol. 5, No.11 (2011).

Strogratz, Steven H. "Exploring Complex Networks". Nature 410 (2011): 268-276.

Von Foerster, Heinz. Las Semillas de la Cibernética. Barcelona: Gedisa, 1991.

Wachhaus, Aroon. "Anarchy as a Model for Network Governance". Public Administration Review 72, No.1 (2011): 33-42.

Werle, Raymund. "The Impact of Information Networks on the Structure of Political Systems". En Understanding the Impact of Global Networks on Local Social, Political and Cultural Values, de C Engel y H. Keller, 159-168. Baden-Baden: Nomos, 1999.

Wilson, Edward O. and Hölldobler, Burt. "Dense Heterarchies and Mass Communication as the Basis of Organization in Ant Colonies". Ecology and Evolutions, Vol. 3, No.3 (1998): 65-68.

Conference Proceedings

Dittrich, Peter, y Lars Winter. "Chemical Organization in a Toy Model of the Political System" European Conference on Complex Systems 2007. Dresden: Complex Systems Society, 2007.

Jones, James Holland, y Mark S. Handcock. "An Assessment of Preferential Attachement as a Mechanism for Human Sexual Network Formation". Proceedings of the Royal Society of Biological Sciences vol. 270, No.1520 (2003): 1123-1128.

Mezza-Garcia, Nathalie. "Bio-Inspired Political Systems: Opening a Field" Proceedings of the European Conference on Complex Systems ECCS´12. Brussels: Springer, 2013.

Nedelcu, Paul-Iulian. "State Structure and Political Regime Structure". International Conference European Integration Realities and Perspectives. 2012. 289-292.

Electronic Documents

Aldana, Maximino. Instituto de Ciencias Físicas de la UNAM, “Redes Complejas.”. November, 2006. Consulted on June, 2012. www.fis.unam.mx/~max/English/notasredes.pdf

Hsu, Sara. "The Effect of Political Regimes on Inequality, 1963-2002, working Paper, University of Texas". September, 2008. Consulted on September, 2012. http://utip.gov.utexas.edu/papers/utip_53.pdf.

Kelly, Charles. "Political Science, Basic Concepts". Kean University. Consulted on October, 2012. http://www.kean.edu/~ckelly/basicconcepts.doc

Mitchell, Melanie. "Complex Systems: Network Thinking". 2006. Consulted on October, 2012. http://www.santafe.edu/media/workingpapers/06-10-036.pdf

United Citied and Local Governments (UCLG). Decentralization and Local Democracy in the World: First Global Report 2008. Barcelona: World Bank Publications, 2008. Consulted on February 2012. http://www.cities-localgovernments.org/gold/Upload/gold_report/01_introduction_en.pdf

Images and Pictures from the web used in tables and figures

"Ants". Consulted on October, 2012. http://www.google.com.co/imgres?hl=es&safe=off&tbo=d&biw=1163&bih=613&tbm=isch&tbnid=6DJ_NdMPFENRSM:&imgrefurl=http://bememachine.blogspot.com/2008/10/superorganism-meme-returns-although-it.html&docid=Qi4RcQrm6eGxWM&imgurl=http://1.bp.blogspot.com/_ZgvU9rSLSG0/SP9Ctl5tZ9I/AAAAAAAAAFQ/CARzl3FViMc/s400/swarmbehavior.jpg&w=324&h=400&ei=hDEHUebqEIra8ASlnoGYBg&zoom=1&ved=1t:3588,r:2,s:0,i:82&iact=rc&dur=804&sig=102864176930550564205&page=1&tbnh=202&tbnw=148&start=0&ndsp=15&tx=93&ty=66

"Bacteria Colony". Consulted on January, 2013. http://eclecticthinktank.wordpress.com/2012/12/16/intelligence-vs-iq-examining-the-foundation-and-future-of-our-educational-system/

"Circulatory System". Consulted on November, 2013. http://www.rogerolivella.net/insula/en/descripcio.htm

"Courtiers Roman Empire". Consulted on October, 2012. http://www.google.com.co/imgres?hl=es&sa=X&tbo=d&biw=1163&bih=613&tbs=itp:lineart&tbm=isch&tbnid=wt9MxnKbzLkxOM:&imgrefurl=http://www.studyblue.com/notes/note/n/exam-1-ch4/deck/2279448&docid=6pOzsEwBkjWGFM&imgurl=http://classconnection.s3.amazonaws.com/190/flashcards/305190/png/a1330345382629.png&w=515&h=864&ei=8lXaUNLGL4vS9ASS-ICwCw&zoom=1&iact=hc&vpx=268&vpy=4&dur=2019&hovh=291&hovw=173&tx=83&ty=41&sig=102864176930550564205&page=4&tbnh=156&tbnw=93&start=91&ndsp=32&ved=1t:429,r:93,s:0,i:367

"Democracy". Consulted on October, 2012. http://www.contrainfo.com/1727/la-inutilidad-manifiesta-de-los-politicos-primera-parte/

"Ecclesia". Consulted on October, 2012. http://www.google.com.co/imgres?start=198&num=10&hl=es&tbo=d&biw=1163&bih=613&tbm=isch&tbnid=Iz6eUapZf7dpgM:&imgrefurl=http://www.heritage-images.com/Preview/PreviewPage.aspx%3Fid%3D1157303%26pricing%3Dtrue%26licenseType%3DRM&docid=ZnzTJ-Wy1FyFtM&imgurl=http://watermarked.heritage-images.com/1157303.jpg&w=435&h=512&ei=bVraUIXeL5H69gTBrYCoBA&zoom=1&iact=hc&vpx=78&vpy=89&dur=959&hovh=244&hovw=208&tx=108&ty=111&sig=102864176930550564205&page=8&tbnh=147&tbnw=125&ndsp=30&ved=1t:429,r:21,s:200,i:67

"Emperor/God". Consulted on October, 2012. http://www.google.com.co/imgres?num=10&hl=es&sa=X&tbo=d&biw=1024&bih=540&tbs=itp:lineart&tbm=isch&tbnid=Ja6wLB48OJmyJM:&imgrefurl=http://www.deadchickenhat.co.uk/kids.php&docid=SD00dzRiW8cS7M&imgurl=http://www.deadchickenhat.co.uk/resources/romeman.gif&w=718&h=957&ei=2zTbULu_NoOe8gT1hIHADQ&zoom=1&iact=hc&vpx=406&vpy=137&dur=1086&hovh=260&hovw=195&tx=87&ty=113&sig=102864176930550564205&page=3&tbnh=141&tbnw=107&start=50&ndsp=29&ved=1t:429,r:75,s:0,i:313

"Fern". Consulted on November, 2013. http://www.rogerolivella.net/insula/en/descripcio.htm

"Feudal lords". Consulted on October, 2012. http://www.google.com.co/imgres?start=314&hl=es&sa=X&tbo=d&biw=1163&bih=613&tbs=itp:lineart&tbm=isch&tbnid=xG9ACvIx8cmGnM:&imgrefurl=http://mcns.blogspot.com/2005_02_01_archive.html&docid=wMepB3ShRY5Z7M&imgurl=http://images.snapfish.com/342%25253B7%25253B9923232%25257Ffp3%25253B%25253Dot%25253E232%25253A%25253D6%25253B%25253C%25253D878%25253DXROQDF%25253E23237775832%25253B2ot1lsi&w=469&h=480&ei=jVbaUKD4GoXe8AT8yYHwAw&zoom=1&iact=hc&vpx=903&vpy=136&

dur=1740&hovh=228&hovw=222&tx=156&ty=105&sig=102864176930550564205&page=11&tbnh=136&tbnw=132&ndsp=32&ved=1t:429,r:21,s:300,i:67

"Fibonacci Fractal". Consulted on January, 2013. http://datachurch.com/category/article/page/19/

"Fish School". Consulted on October, 2012. http://aquariumprosmn.com/2010/01/460/

"Flock of starlings". Consulted on October, 2012. http://www.google.com.co/imgres?imgurl=http://pathtothepossible.files.wordpress.com/2011/10/starlings.jpg&imgrefurl=http://pathtothepossible.wordpress.com/tag/starlings/&usg=__XqtmjIjNcQgAd5x916310TBcmrw=&h=430&w=500&sz=120&hl=es&start=0&sig2=BYkhrtW7NtiryjfNZEwhvg&zoom=1&tbnid=dkRVNxoEMrUKwM:&tbnh=190&tbnw=242&ei=FUAJUeauDISu9ATEjoCgCQ&itbs=1&ved=1t:3588,r:0,s:0,i:82&iact=rc&dur=662&sig=102864176930550564205&page=1&ndsp=16&tx=72&ty=39

"Fractal Fern". Consulted on November, 2013. http://www.tattoodonkey.com/fern-leaf/spaennare.se*FRACTAL*fern.jpg/http://www.google.com/imgres?imgurl=http://datachurch.com/wp-content/uploads/2009/08/fractal-280x300.png&imgrefurl=http://datachurch.com/category/article/page/19/&h=300&w=280&sz=52&tbnid=ap7EVJZAU6JqKM&tbnh=232&tbnw=217&zoom=1&usg=__ArqCY0_RbnLTpe3EUkw8O6gN-vc=&hl=en&sa=X&ei=dMYuUdKlLIyk8ASE5oCICw&ved=0CCwQ8g0

"Galaxy M81". Consulted on October, 2012. https://www.sflorg.com/observatories/spitzer/spitzer_04?full=1

"Imperial guard". Consulted on October, 2012.
http://www.comp.dit.ie/dgordon/lectures/hum1/031020/031020hum.htm

“Iterated Sierpinski”. Consulted on January, 2013. http://portaldoprofessor.mec.gov.br/fichaTecnicaAula.html?aula=22040

“Julia Set”. Consulted on January, 2013. http://www.math.harvard.edu/~jjchen/fractals/

“Justice”. Consulted on October, 2012. http://www.google.com.co/imgres?hl=es&tbo=d&biw=1024&bih=540&tbs=itp:lineart&tbm=isch&tbnid=J6PbBabNMdhqsM:&imgrefurl=http://www.canstockphoto.com/illustration/justice.html&docid=cCf_ytqJAapVzM&imgurl=http://ec.l.thumbs.canstockphoto.com/canstock5372255.jpg&w=147&h=150&ei=6lLbUOvrMIP-9QTnioHYDA&zoom=1&iact=hc&vpx=193&vpy=12&dur=818&hovh=120&hovw=117&tx=68&ty=86&sig=102864176930550564205&page=4&tbnh=120&tbnw=117&start=68&ndsp=28&ved=1t:429,r:90,s:0,i:385

“Koch Snowflake”. Consulted on December, 2013. http://www.tumblr.com/tagged/koch%20snowflake

“Lightning”. Consulted on November, 2012. http://www.rogerolivella.net/insula/en/descripcio.htm

“Magistrates”. Consulted on October, 2012. http://www.bible-history.com/archaeology/rome/legion-camp.html

“Mandelbrot Set”. Consulted on December, 2013. http://escara-fotoprojekte.blogspot.com/2010_11_01_archive.html

“River with Tributaries”. Consulted on November, 2013. http://zonavip.net/imagenes/14998242/parque-nacional-y-natural-de-donana-armonia-fr.html

"Roman Crown". Consulted on October, 2012. http://www.google.com.co/imgres?num=10&hl=es&sa=X&tbo=d&biw=1163&bih=613&tbs=itp:clipart&tbm=isch&tbnid=aaaFT4TqsYjYiM:&imgrefurl=http://forum.playdom.com/showthread.php%3F58503-A-Civic-Crown-to-wear-together-with-my-toga!&docid=oh1aV7NiiijAAM&imgurl=http://image.spreadshirt.com/image-server/v1/compositions/15693998/views/1,width%253D178,height%253D178/laurel-wreath-roman-caesar-1c_design.png&w=178&h=178&ei=u2baUNnqCIL88QSDyoHYAQ&zoom=1&iact=hc&vpx=4&vpy=4&dur=570&hovh=142&hovw=142&tx=88&ty=79&sig=102864176930550564205&page=1&tbnh=139&tbnw=139&start=0&ndsp=24&ved=1t:429,r:0,s:0,i:83

"President". Consulted on October, 2012. http://www.google.com.co/imgres?hl=es&sa=X&tbo=d&biw=1163&bih=613&tbs=itp:lineart&tbm=isch&tbnid=d8H71UV6p8ZvqM:&imgrefurl=http://es.123rf.com/imagenes-de-archivo/autoritario.html&docid=NpHJOsR69SHP8M&imgurl=http://us.cdn3.123rf.com/168nwm/dmiskv/dmiskv1210/dmiskv121000107/15652253-apuntando-el-dedo.jpg&w=168&h=168&ei=8V_aUPuQLoem8ATe7YC4DA&zoom=1&iact=hc&vpx=4&vpy=192&dur=964&hovh=134&hovw=134&tx=81&ty=55&sig=102864176930550564205&page=1&tbnh=131&tbnw=123&start=0&ndsp=23&ved=1t:429,r:0,s:0,i:83

"Prime Minister". Consulted on October, 2012. http://www.google.com.co/imgres?hl=es&tbo=d&biw=1163&bih=613&tbs=ic:specific,isc:black,itp:clipart&tbm=isch&tbnid=mJg3A3dWCCvR5M:&imgrefurl=http://www.trevorloudon.com/tag/occupy-wall-street/&docid=Ro1DN5Qc-iZqgM&imgurl=http://www.trevorloudon.com/wp-content/uploads/2012/03/day6.png&w=262&h=238&ei=rOnZUIOhEoi69gT2hIC

QCg&zoom=1&iact=hc&vpx=618&vpy=129&dur=1490&hovh=190&hovw=210&tx=86&ty=94&sig=102864176930550564205&page=2&tbnh=123&tbnw=135&start=20&ndsp=30&ved=1t:429,r:33,s:0,i:199

"Roman Republic". Consulted on October, 2012. http://www.mariamilani.com/rome_pictures/Roman_Society_Republic_vote.htm

"Romanescu". Consulted on December, 2013. www.google.com.co/imgres?imgurl=http://onlyhdwallpapers.com/thumbnail-small/fractals_fibonacci_broccoli_desktop_2896x1944_hd-wallpaper-538466.jpg&imgrefurl=http://onlyhdwallpapers.com/high-definition-wallpaper/broccoli-desktop-hd-wallpaper-439972/&h=135&w=180&sz=7&tbnid=wN46Tn6ih2sCWM:&tbnh=91&tbnw=121&zoom=1&usg=__uivVhKvDNPZ1PZ3G-cFrFkddZyY=&docid=D47OPyMX6gkJrM&hl=es&sa=X&ei=Ts4uUcvsJIai8ASGnIH4AQ&ved=0CC8Q9QEwAQ&dur=475

"Sierfs". Consulted on October, 2012. http://www.forumromanum.org/life/johnston_9.html

"Snowflake". Consulted on January, 2013. http://www.lakesleisure.co.nz/news/N191-Sarahs-Splash-Blog---Inflatable-Fun-Technical-Inspiration-Pool-Closures/

"Spartans". Consulted on October, 2012. http://www.google.com.co/imgres?hl=es&sa=X&tbo=d&biw=1163&bih=613&tbs=ic:specific,isc:black,itp:clipart&tbm=isch&tbnid=HkTAvO6-qpqb3M:&imgrefurl=http://es.123rf.com/imagenes-de-archivo/espartano.html&docid=KDEvW6YoyxBrUM&imgurl=http://us.cdn3.123rf.com/168nwm/alextrim/alextrim1009/alextrim100900003/7831937-el-soldado-griego-esta-dispuesto-a-defender-la-libertad-en-lucha-desigual.jpg&w=74&h=168&ei=feXZUPLKDpGI9gSRrID4Cg&zoom=1&iact=hc&vpx=965&vpy=348&dur=1369&hovh=134&hovw=60&tx=80&ty=42&sig=10

286417693055064205&page=1&tbnh=134&tbnw=60&start=0&ndsp=24&ved=1t:429,r:15,s:0,i:140

"Spartan Clothes". Consulted on October, 2012. http://karenswhimsy.com/ancient-greek-clothing.shtm

"The people, Roman Empire". Consulted on October, 2012. http://www.deadchickenhat.co.uk/resources/romewoman.gif

"The people: Feudalism". Consulted on October, 2012. http://www.google.com.co/imgres?num=10&hl=es&sa=X&tbo=d&biw=1024&bih=540&tbs=itp:lineart&tbm=isch&tbnid=gXLNup8g_vMsGM:&imgrefurl=http://www.oncoloring.com/middle-ages-coloring-pages.html&docid=BMtuOuXdvZEzvM&imgurl=http://img.oncoloring.com/young-farmers-living-unde_49f8137692bf0-p.gif&w=280&h=262&ei=BGnbUJ-ROoGm8QSty4HAAQ&zoom=1

"Top of a Tree". Consulted on January, 2013. http://www.rogerolivella.net/insula/en/descripcio.htm